\documentclass[a4paper, aps, prb, onecolumnm, secnumarabic, 12pt]{article}
\usepackage{mathtools}
\usepackage{units}
\usepackage{subcaption} 
\usepackage{titling}
\usepackage{hyperref}
\usepackage{verbatim}
\usepackage{graphicx}
\usepackage{wrapfig}
\usepackage{sectsty}
\sectionfont{\fontsize{12}{15}\selectfont}
\usepackage{tocloft}

\usepackage[onehalfspacing]{setspace}
\usepackage[sort&compress,numbers]{natbib}
\title{Time- and spatially-resolved magnetization dynamics driven by spin-orbit torques}
\author{M. Baumgartner,$^{1}$ K. Garello,$^{1,2}$ J. Mendil,$^{1}$ C. O. Avci,$^{1}$ E. Grimaldi,$^{1}$\\ C. Murer,$^{1}$ J. Feng,$^{1}$ M. Gabureac,$^{1}$ C. Stamm,$^{1}$ Y. Acremann,$^{3}$\\S. Finizio,$^{4}$ S. Wintz,$^{4}$ J. Raabe,$^{4}$ P. Gambardella$^{1}$\\
\\
\normalsize{$^{1}$Department of Materials, ETH Z\"urich, 8093 Z\"urich, Switzerland,}\\
\normalsize{$^{2}$IMEC, Kapeldreef 75, 3001 Leuven, Belgium,}\\
\normalsize{$^{3}$Laboratory for Solid State Physics, ETH Z\"urich, 8093 Z\"urich, Switzerland,}\\
\normalsize{$^{4}$Paul Scherrer Institute, 5232 Villigen PSI, Switzerland}}
\date{}

\begin{document}

\begin{titlingpage}
	\maketitle
	\begin{abstract}
Current-induced spin-orbit torques (SOTs) represent one of the most effective ways to manipulate the magnetization in spintronic devices. The orthogonal torque-magnetization geometry, the strong damping, and the large domain wall velocities inherent to materials with strong spin-orbit coupling make SOTs especially appealing for fast switching applications in nonvolatile memory and logic units. So far, however, the timescale and evolution of the magnetization during the switching process have remained undetected. Here, we report the direct observation of SOT-driven magnetization dynamics in Pt/Co/AlO$_x$ dots during current pulse injection. Time-resolved x-ray images with 25~nm spatial and 100~ps temporal resolution reveal that switching is achieved within the duration of a sub-ns current pulse by the fast nucleation of an inverted domain at the edge of the dot and propagation of a tilted domain wall across the dot. The nucleation point is deterministic and alternates between the four dot quadrants depending on the sign of the magnetization, current, and external field. Our measurements reveal how the magnetic symmetry is broken by the concerted action of both damping-like and field-like SOT and show that reproducible switching events can be obtained for over 10$^{12}$ reversal cycles.
	\end{abstract}
\end{titlingpage}

Controlling the magnetic state of ultrathin heterostructures using electric currents is key to developing nonvolatile memory devices with minimal static and dynamic power consumption\cite{Kent2015}. A promising approach for magnetic switching is based on injecting an in-plane current into a ferromagnet/heavy metal bilayer, where the spin-orbit torques (SOTs)\cite{Garello2013,Kim2013} resulting from the spin Hall effect and interface charge-spin conversion\cite{Sinova2015,Manchon2008,Freimuth2015a,Wang2016,Stiles2016b} provide an efficient mechanism to 
reverse the magnetization\cite{Miron2011,Liu2012,Garello2014,Fukami2016b,Ghosh2017,Yu2014b,Safeer2016} and manipulate domain walls (DWs)\cite{Miron2011b,Emori2013,Ryu2013,Haazen2013}.

SOT switching schemes can be easily integrated into three-terminal magnetic tunnel junctions having either in-plane\cite{Liu2012} or out-of-plane\cite{Cubukcu2014} magnetization. Although the three-terminal geometry is more demanding in terms of size, 
it offers desirable features compared to the two-terminal spin-transfer torque (STT) approach presently used in magnetic random access memories (MRAM) \cite{Prenat2016}. One such feature is the separation of the read and write current paths in the tunnel junction, which avoids electrical stress of the oxide barrier during writing and allows for independent optimization of the tunneling magnetoresistance during reading. 
The other crucial feature is the switching speed, which is expected to be extremely fast because the spin accumulation inducing the SOTs is orthogonal to the quiescent magnetization, leading to a negligible incubation delay. Such a delay is a major issue for STT devices, since thermal fluctuations result in a switching time distribution that is several ns wide\cite{Devolder2008,Hahn2016}. Furthermore, the SOT-induced magnetization dynamics is governed by strong damping in the monodomain regime\cite{Lee2013,Park2014} and fast domain wall motion in the multidomain regime\cite{Miron2011b,Emori2013,Ryu2013}, both favoring rapid reversal of the magnetization.

Recent investigations of SOTs in ferromagnet/heavy metal layers showed that reliable switching can be achieved by the injection of current pulses as short as 200~ps in Pt/Co and Ta/CoFeB structures \cite{Garello2014,Zhang2015c,Aradhya2016}.
However, these experiments only measured the switching probability as a function of pulse amplitude and duration, while the mechanism and the timescale of magnetization reversal remain unknown. Microscopy studies performed using the magneto-optic Kerr effect have extensively probed SOT-induced DW 
displacements\cite{Yu2014b,Safeer2016,Miron2011b,Emori2013,Ryu2013,Haazen2013,LoConte2014}, revealing the role played by the Dzyaloshinskii-Moriya interaction (DMI) in stabilizing chiral DW structures that have very high mobility\cite{Emori2013,Ryu2013,Thiaville2012,Boulle2013,Martinez2013,Perez2014,Martinez2014}. Such investigations have a spatial resolution of the order of 1~$\mu$m, but only probed the static magnetization after current injection, similar to the pulsed switching experiments\cite{Garello2014,Zhang2015c,Aradhya2016}.
In parallel to this activity, several theoretical models have been proposed in order to elucidate the SOT-induced dynamics. The most straightforward approach is based on the macrospin approximation\cite{Lee2013,Lee2014a,Park2014,Legrand2015}, which applies in the limit of small magnets and coherent rotation of the magnetization. Under these assumptions, the damping-like (DL) torque 
$\mathbf{T}^{DL} \propto \mathbf{M}\times (\mathbf{y}\times \mathbf{M})$ induces the rotation of the magnetization ($\mathbf{M}$) while the field-like (FL) torque $\mathbf{T}^{FL} \propto \mathbf{M}\times \mathbf{y}$ favors precessional motion of the magnetization about the $y$-axis, orthogonal to the current. Such a model is often used to relate the critical switching current to the SOT amplitude in practical devices\cite{Zhang2015c,Fukami2016b}, even though the reversal mode is still a matter of debate. A second model is based on the random nucleation and isotropic expansion of magnetic bubbles, induced by a thermally-assisted DW depinning process and the DL component of the SOT\cite{Lee2014}.
Finally, micromagnetic models have been proposed, in which domain nucleation is either random and thermally-assisted\cite{Finocchio2013,Perez2014} or determined by the competition between external field and DMI at the sample edge followed by DW propagation across the magnetic layer\cite{Mikuszeit2015,Martinez2015}. In such schemes, which apply to perpendicularly magnetized layers, the DL torque acts as a $z$-axis field $B^{DL} \propto M_x$ on the internal DW magnetization, favoring the expansion of a domain on the side where $|M_x|$ is larger (for example at an up-down DW of the type $\uparrow \leftarrow \downarrow$ rather than at the opposite down-up DW $\downarrow \rightarrow\uparrow$).

Despite these considerable efforts, a number of issues remain outstanding. The most prominent concern the actual timescale and physical processes that govern the magnetization dynamics during and after current injection, the initial and intermediate magnetic configuration preceding switching, as well as the role played by the DL and FL components of the torque and the DMI. 
Here, we report the direct observation of the dynamics excited by SOTs during the switching process by performing current-pump and x-ray probe experiments in the time domain using a scanning transmission x-ray microscope (STXM)\cite{Raabe2008}.
Our measurements reveal the evolution of the nanoscale magnetization from the initial domain nucleation stage to full or partial switching, depending on the duration and amplitude of the current pulses. These results determine the actual speed of the switching process and show how the reversal path is unique to SOTs and predetermined by the interplay of the current-induced torques, DMI, and external magnetic field. 
We find four different configurations of the nucleation point and fast direction of DW propagation, corresponding to the four possible combinations of current and field polarization. The STXM data, combined with micromagnetic simulations and all-electrical measurements of the switching probability, provide a consistent picture of the switching process and evidence how the FL torque influences the reversal path together with the DL torque and DMI. Finally, we show that switching is robust with respect to repeated cycling events as well as to the presence of defects in the sample magnetic structure, which is very promising for the application of SOTs in MRAM devices.

%
\begin{figure}[t!]
	\centering
	\includegraphics[width=16cm]{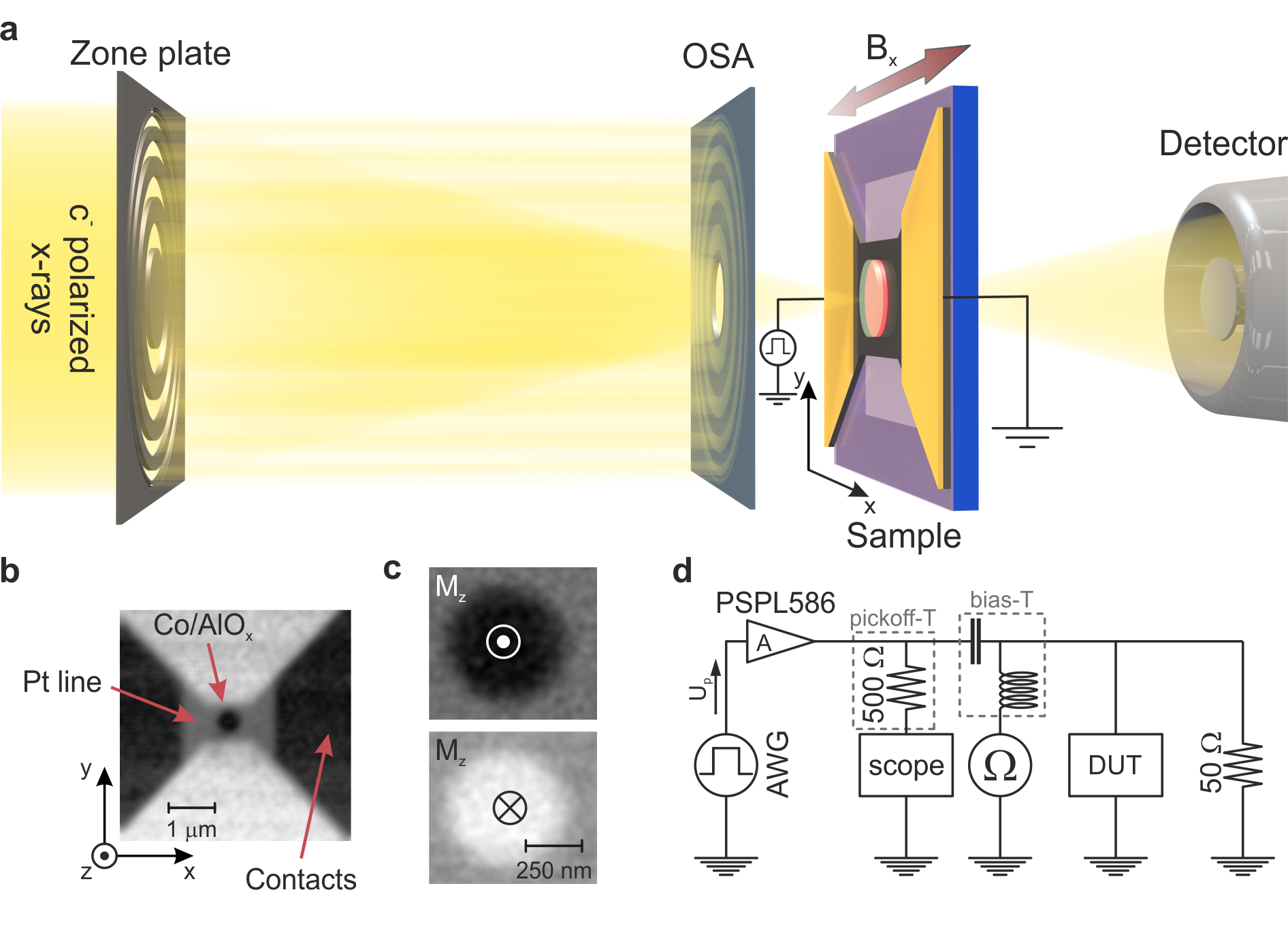}
	\caption{\textbf{Schematic of the experimental setup and sample structure.} \textbf{a,} A circularly polarized x-ray beam is focused to a 25~nm spot by a zone plate and an order selecting aperture (OSA). The transmission of the x-rays through the sample is monitored by an avalanche photodiode as the sample stage is scanned in steps of \unit[12-25]{nm} across the x-ray focus. \textbf{b,} Representative STXM image of a Co dot and Pt line showing topographic and elemental contrast at a photon energy of 779~eV. The darker regions indicate the position of the Au contact pads and of the Co dot, which absorb x-rays more strongly than the bare Pt and Si$_3$N$_4$ regions. \textbf{c,} Magnetic contrast due to the XMCD effect: magnetization pointing parallel or anti-parallel to the x-ray helicity results in different absorption intensity. Black (white) contrast indicates $M_z > 0$ ($M_z < 0$). \textbf{d,} Scheme of the pump current circuit.}
	\label{setup:pic}
\end{figure}
\paragraph*{Time-resolved switching}
We studied circular-shaped Co dots with perpendicular magnetization that are 1~nm thick and 500~nm in diameter, capped by 2~nm AlO$_x$ and deposited on a 750~nm wide, 5~nm thick Pt current line. The samples were fabricated using electron-beam lithography on transparent Si$_3$N$_4$ membranes suitable for x-ray transmission and contacted by Au leads, as illustrated in Fig.~\ref{setup:pic}c. Replicas of the dots with four electrical contacts were fabricated on Si$_3$N$_4$ membranes for all-electrical pulsed switching together with Hall bar structures for SOT measurements (see Supplementary Information). Figure~\ref{setup:pic}b shows a static image of the Co dot and Pt line obtained by scanning the sample under the focused x-ray beam. Time-resolved images of the Co magnetization were recorded stroboscopically by probing the transmitted x-ray intensity at the Co $L_{3}$ absorption edge, where x-ray magnetic circular dichroism (XMCD) provides contrast to the out-of-plane component of the magnetization ($M_z$), as shown in Fig.~\ref{setup:pic}c. The pump current consists of a sequence of negative (set) and positive (reset) pulses of variable length and amplitude with a period of 102.1~ns, corresponding to a switching rate of about $\unit[20]{MHz}$.
An in-plane magnetic field $B_x$ was applied along the current direction in order to uniquely define the switching polarity\cite{Miron2011}.

%
\begin{figure}
	\centering
	\includegraphics[width=16cm]{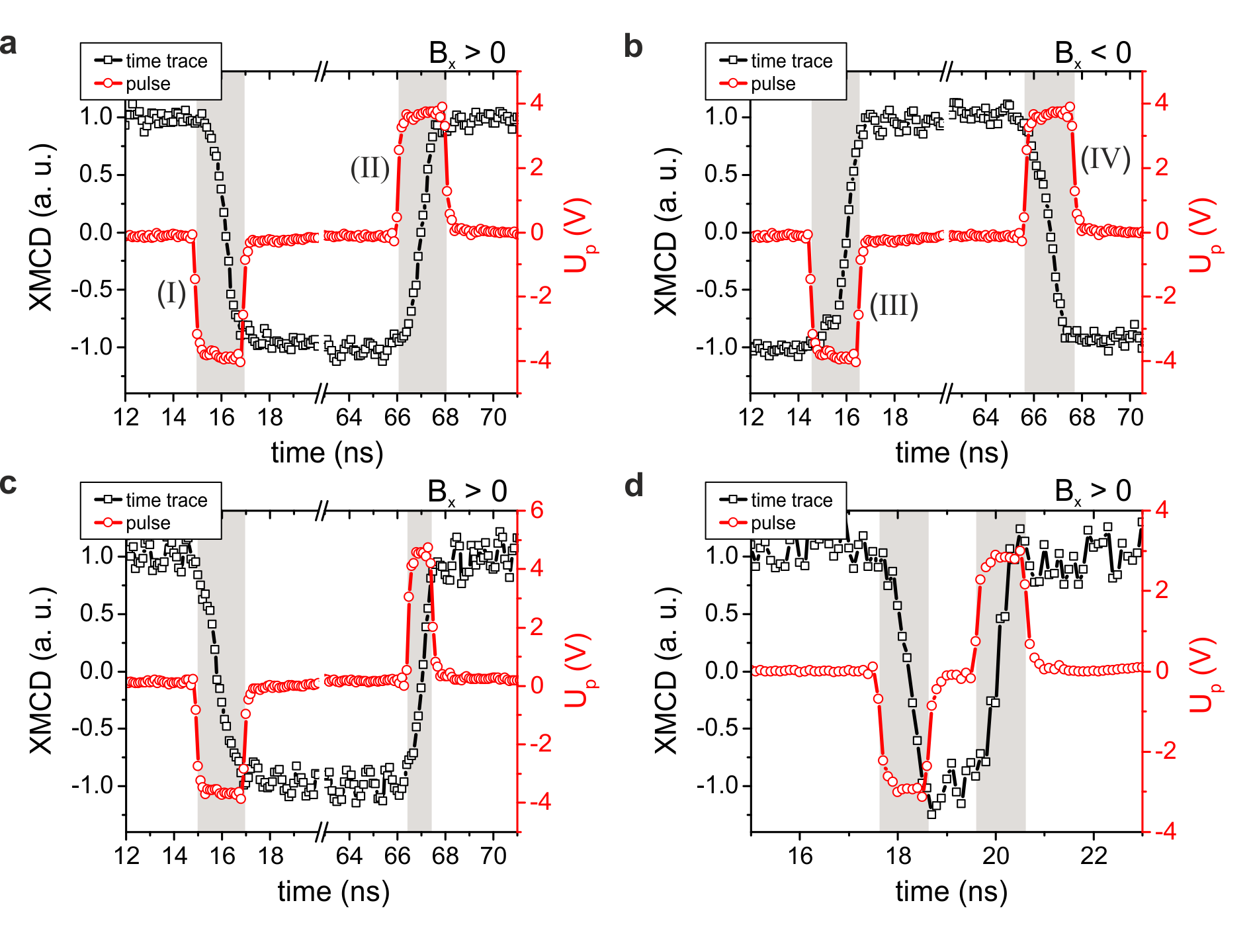}
	\caption{\textbf{Time-resolved magnetization switching.} XMCD time traces showing reversal of the magnetization during the injection of \unit[2]{ns} long current pulses of opposite polarity at \textbf{a,} $B_x = \unit[94]{mT}$ and \textbf{b,} $B_x = \unit[-124]{mT}$. The sign of the XMCD signal is positive for $M_z < 0$ and negative for $M_z > 0$. \textbf{c,} Asymmetric set-reset sequence in which the first and second pulses are 2~ns and \unit[0.8]{ns} long, respectively. \textbf{d,} Fast toggle sequence consisting of two \unit[1]{ns} long current pulses of opposite polarity separated by \unit[1]{ns} with in-plane field $B_x = \unit[92]{mT}$. $U_p =\unit[1]{V}$ corresponds to a current density $j_p = \unit[8.4\cdot10^7]{A/cm^2}$ in the Pt line.}
	\label{timeTraces:pic}
\end{figure}
Figure~\ref{timeTraces:pic}a shows the XMCD time trace obtained by integrating the transmitted x-ray intensity over the entire area of the Co dot during the current pulse sequence, which was separately recorded by a fast oscilloscope in parallel with the sample. The pulses in this sequence are \unit[2]{ns} long, with a rise time of about \unit[150]{ps}, and have a separation of 50~ns. The XMCD time trace shows that the magnetization switches from down to up during the first (negative) pulse and from up to down during the second (positive) pulse.
The timescale over which $M_z$ changes between two states with opposite saturation magnetization coincides with the duration of the current pulses, indicating that the magnetization is fully reversed within the pulse interval, with no significant delay or after-pulse relaxation.
The response of the magnetization to the current pulses inverts upon changing the sign of $B_x$, as shown in Fig.~\ref{timeTraces:pic}b, consistently with previous reports of SOT-induced switching of perpendicularly magnetized layers\cite{Miron2011,Liu2012,Garello2014,Fukami2016b,Yu2014b,Ghosh2017,Emori2013}. Further, we find that the switching speed increases by increasing the current amplitude, allowing for full magnetization reversal within a sub-ns current pulse (Fig.~\ref{timeTraces:pic}c). This speed, and the absence of after-effects, enable the realization of very fast magnetic-writing cycles. One such cycle is shown in Fig.~\ref{timeTraces:pic}d, where we toggle the magnetization between up and down states using a 1~ns on $|$ 1~ns off $|$ 1~ns on pulse sequence. Remarkably, our samples survive uninterrupted pulse sequences for hours without appreciable magnetic or electrical degradation. We tested more than $10^{12}$ consecutive and successful switching events, at current densities of the order of $\unit[2-4\cdot10^8]{A/cm^2}$. Thus, the combination of speed and endurance revealed by our measurements is extremely promising for the operation of SOT-based MRAM and logic devices.

%
\begin{figure}[t!]
	\centering
	\includegraphics[width=16cm]{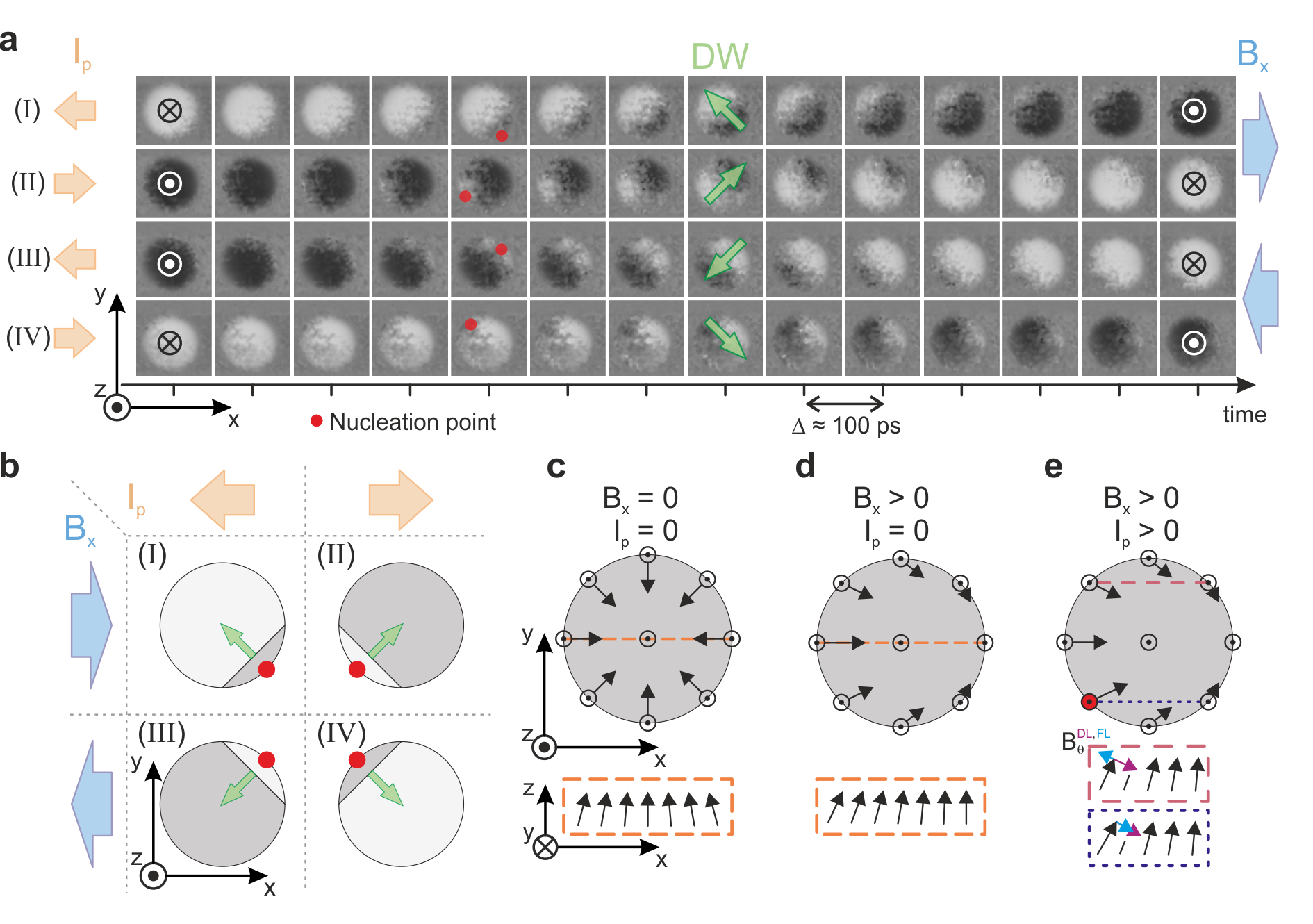}
	\caption{\textbf{Evolution of the magnetization during the switching process.} \textbf{a,} Images taken at intervals of 100~ps during the injection of 2~ns long current pulses. Rows (I,II) and (III,IV) correspond to the time traces shown in Fig.~\ref{timeTraces:pic}a and b, respectively. The red dots indicate the DW nucleation point and the green arrows its propagation direction. The images are low-pass filtered for better contrast (see the Supplementary Information for the raw data and movies). \textbf{b,} Schematic representation of the observed DW nucleation and propagation geometry. Illustration of the DW nucleation process corresponding to case (II): \textbf{c,} Canting of the magnetization at the dot edges induced by the DMI. \textbf{d,} Breaking of the canting symmetry induced by $B_x$. \textbf{e,} Action of $B^{DL}_{\theta}$ and $B^{FL}_{\theta}$.}
	\label{DMI-switching:pic}
\end{figure}
\paragraph*{Spatial evolution of the magnetization during current injection}
We next focus on the transient magnetic configurations and the mechanisms leading to magnetization reversal. Figure~\ref{DMI-switching:pic}a shows four series of consecutive images recorded at time intervals of 100~ps during the switching process, corresponding to the four possible combinations of current and field polarity.
The magnetization reverses by domain nucleation and propagation in all cases, with no appreciable incubation delay. Although the stroboscopic character of our measurements does not allow us to investigate stochastic effects, the observation of a clear DW front moving from a fixed nucleation point on one side of the sample to the opposite side, as shown by the red dots and green arrows in Fig.~\ref{DMI-switching:pic}a, indicates that the reversal process is reproducible and deterministic. Furthermore, as we argue in the following, such a reversal scheme is unique to SOTs. Our data show that domain nucleation takes place at the edge of the sample where the DMI and $B_x$ concur to tilt the magnetization towards the current direction,
thereby confirming the prediction of recent micromagnetic models\cite{Mikuszeit2015,Martinez2015}.
However, the concerted action of $B^{DL}$, DMI, and $B_x$ only leads to a left/right asymmetry, similar to the asymmetric nucleation induced by a perpendicular magnetic field\cite{Pizzini2014}, whereas we observe that the domain nucleation site alternates between the four quadrants shown in Fig.~\ref{DMI-switching:pic}b, with an additional top/bottom asymmetry.

\paragraph*{Edge nucleation}
To explain this additional asymmetry, we have to analyze the effects of the static and dynamic fields in more detail. Pt/Co/AlO$_x$ layers have positive DMI, which stabilizes left-handed N\'eel-DW\cite{Belmeguenai2015} and induces canting of the magnetization at the edge of the dot. In the absence of current and external field, the magnetic moments are symmetrically canted inwards (outwards) of the dot for $M_z > 0$ ($M_z < 0$), as illustrated in Fig.~\ref{DMI-switching:pic}c. Upon applying a static field $B_x$, the canting angle increases on one side, while it decreases on the other (Fig.~\ref{DMI-switching:pic}d), favoring domain nucleation on the side where the canting is larger. The injection of a positive current pulse generates an effective DL field with a polar component $B^{DL}_{\theta}$ that points toward the current direction (purple arrow in Fig.~\ref{DMI-switching:pic}e). For a positive $B_x$, $B^{DL}_{\theta}$ thus leads to left nucleation if $M_z > 0$ and no nucleation if $M_z < 0$. A similar reasoning is valid for all four combinations of field and current polarity leading to magnetization switching, which explains the left/right asymmetry observed in our data.
The top/bottom asymmetry, however, can be explained only if an additional torque plays a significant role in the nucleation process. In the following, we argue that the FL torque, combined with the DMI-induced canting at the sample edges, accounts for such an asymmetry. Because the corresponding effective field $B^{FL}$ points along the $y$-direction and has no projection along the easy axis, the effects of this torque are usually neglected in models of the SOT-induced DW dynamics. However, the polar component $B^{FL}_{\theta} \propto M_zM_y/\sin\theta$ points upward or downward depending on the sign of $M_y$, as illustrated by the blue arrows in Fig.~\ref{DMI-switching:pic}e. Therefore, the rotation of the magnetization and the nucleation of a reverse domain are favored whenever $B^{FL}_{\theta}$ and $B^{DL}_{\theta}$ are parallel to each other, as indicated by the red dots in Fig.~\ref{DMI-switching:pic}, and hindered when they are antiparallel. This qualitative argument is supported by harmonic Hall voltage measurements\cite{Garello2013} of the FL and DL torques in our samples, which show that $B^{FL}$ and $B^{DL}$ have comparable amplitudes of about \unit[20]{mT} per $\unit[10^{8}]{A/cm^2}$, and by both macrospin and micromagnetic simulations that take these torques into account (see Supplementary Information). We note that, in principle, also the $z$-component of the Oersted field generated by the current flowing in the Pt line can induce a similar top/bottom asymmetry as reported here and assist the nucleation process. However, while Oersted field-assisted switching is of interest for device applications, we find that its effects are minor in the present case, as the closest edge of the Co dot is about 125 nm from the edge of the current line and the Oersted field is significantly smaller compared to the FL torque (see Supplementary Information).

\paragraph*{Role of the FL torque in promoting the switching efficiency}

\begin{figure}
	\centering
	\includegraphics[width=16cm]{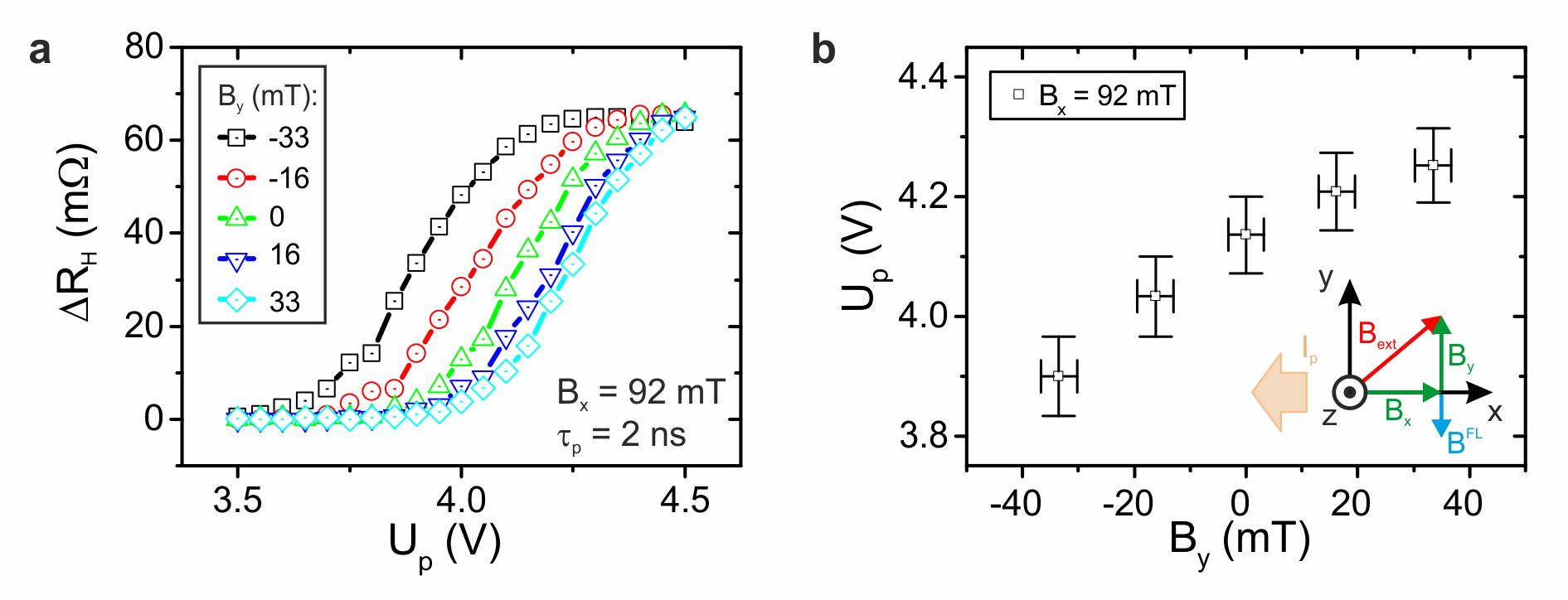}
	\caption{\textbf{Effect of an external field balancing the FL torque.} \textbf{a,} Magnetization change measured by recording the difference of the Hall resistance before and after the pulse as a function of the applied voltage $U_p$ and the in-plane external field $B_y$. $\Delta R_H$ is proportional to the fraction of the dot area that has reversed its magnetization, averaged over 200 switching events. The $x$-component of the external field is fixed to $\unit[92]{mT}$. \textbf{b,} Threshold voltage at 50~\% switching as a function of $B_y$. Inset: schematic of the field direction. For a negative current, $B_y$ opposes (favors) $B^{FL}$ for $B_y > 0$ ($B_y < 0$).}
\label{TransportFLdependence:pic}
\end{figure}

To further investigate the effect of the FL torque on the reversal process, we performed all-electrical pulsed switching experiments on the replica Pt/Co/AlO$_x$ dots in the presence of an additional in-plane field $B_y$, applied parallel or antiparallel to $B^{FL}$. In these experiments, the magnetization of the dot is measured by the anomalous Hall effect after the injection of each current pulse.
Figure~\ref{TransportFLdependence:pic}a shows the change of the Hall resistance measured after the injection of negative pulses of increasing amplitude for different values of $B_y$. The shift of the different curves, exemplified by the voltage threshold at which 50 \% of the dot has reversed its magnetization (Fig.~\ref{TransportFLdependence:pic}b), shows that a negative $B_y$, which is parallel to $B^{FL}$ and antiparallel to the in-plane component of the Oersted field, promotes switching, whereas a positive $B_y$ hampers it. These data support the conclusions drawn from the analysis of the nucleation point, namely that the FL torque plays a crucial role in triggering the reversal process and in enhancing the switching efficiency.

\begin{figure}[t!]
	\centering
	\includegraphics[width=16cm]{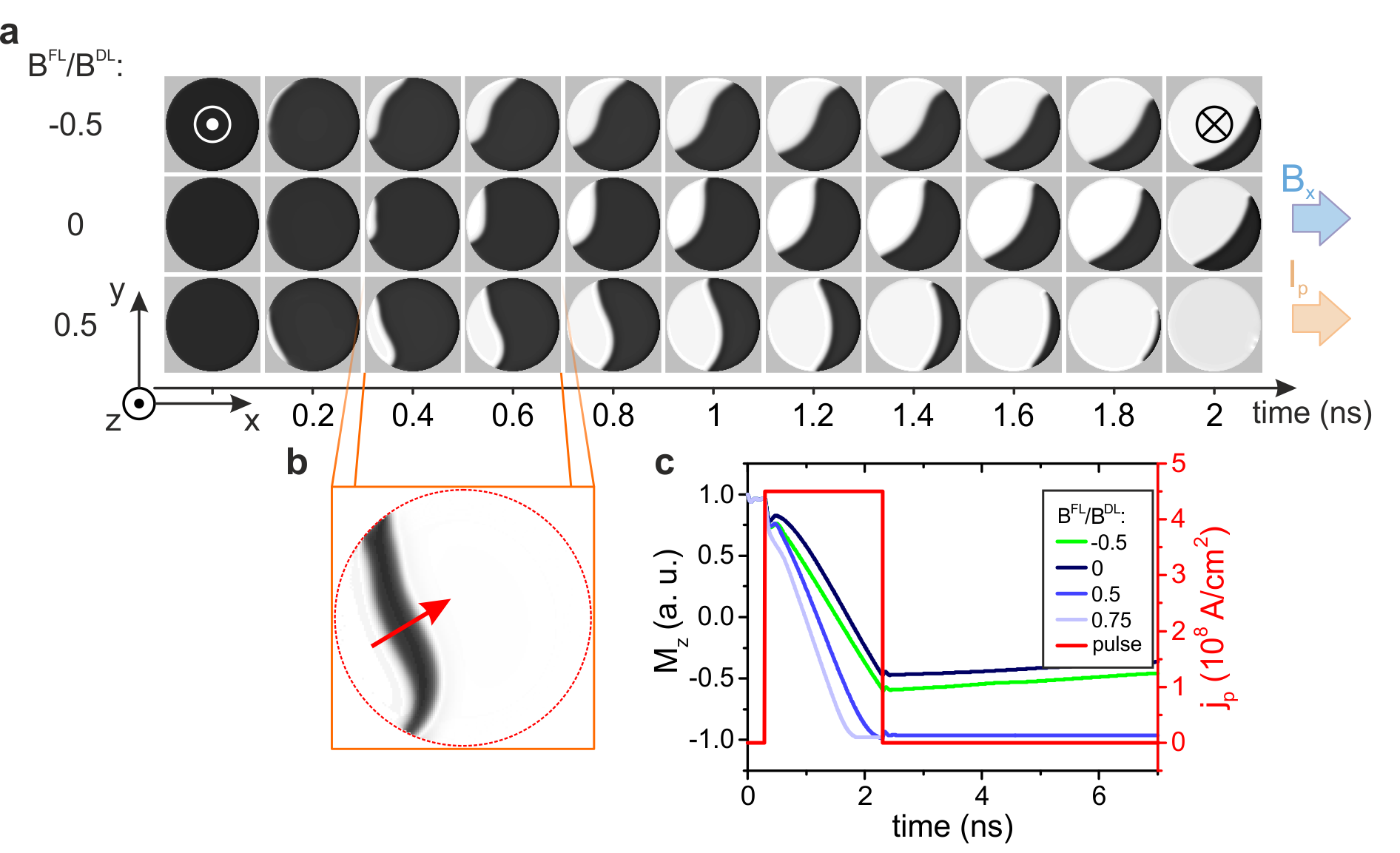}
	\caption{\textbf{Micromagnetic simulations of the reversal process.} \textbf{a,} Snapshots of the magnetic configuration at different times. The simulations are run for different values of the FL torque relative to the DL torque (see Methods). \textbf{b,} Contrast difference between two consecutive frames, showing the tilt of the DW. The black area represents the expansion of the reversed domain. \textbf{c,} Simulated time trace of the perpendicular magnetization component during switching.}
	\label{FL-simulations}
\end{figure}

\paragraph*{Dynamic domain wall propagation}
A striking feature observed in Figure~\ref{DMI-switching:pic} is that the propagating DW front is tilted relative to the current direction, with a tilt angle of about $45^{\circ}$ that changes in steps of $90^{\circ}$ depending on the up/down or down/up DW configuration and the sign of the current. According to recent studies, current-induced DW tilting is a telltale signature of the DMI in perpendicular magnetized nanotracks\cite{Boulle2013,Martinez2014}. However, the tilt angle in Fig.~\ref{DMI-switching:pic} is opposite to that predicted by micromagnetic models of Pt/Co heterostructures\cite{Boulle2013,Martinez2014,Martinez2015} and observed by MOKE microscopy in Pt/Co/Ni/Co racetracks\cite{Ryu2012}. More specifically, the angle between the DW normal and the current direction is $\approx -45^{\circ}$ for a left-handed up-down DW ($\uparrow\leftarrow\downarrow$) at positive current (see panel IV in Fig.~\ref{DMI-switching:pic}b) rather than $\approx +45^{\circ}$ as reported in previous studies. We believe that this inconsistency stems from the neglect of the FL torque in the micromagnetic models of current-induced DW motion as well as from the time-resolved nature of our measurements. The tilt angle in the Pt/Co/AlO$_x$ dots is in fact analogous to that induced by an external in-plane magnetic field $B_y$, which leads to a rotation of the internal DW magnetization away from the $x$-axis in order to recover the N\'{e}el configuration favored by the DMI\cite{Boulle2013,Martinez2014}. Our micromagnetic simulations, which include $B^{FL}$ as well as $B^{DL}$ and the DMI, correctly reproduce the observed dynamic tilt during DW propagation, as shown in Fig.~\ref{FL-simulations}. Moreover, the simulations indicate that the FL torque promotes faster DW propagation in the direction indicated by the arrow in Fig.~\ref{FL-simulations}b, which coincides with case II reported in Fig.~\ref{DMI-switching:pic} and may also explain the strong anisotropy of the DW velocity recently reported in extended Pt/Co/AlO$_x$ layers\cite{Safeer2016}. We note further that the DW propagation direction is not related to the bias field $B_x$, the main purpose of which is to break the spin-canting symmetry due to the DMI, whereas the DW velocity increases with $B_x$ (see Supplementary Information). As the tilt angle depends on the $B^{FL}/B^{DL}$ ratio, the opposite DW tilt relative to the Pt/Co/Ni/Co racetracks\cite{Ryu2012} may be explained by the different SOT amplitudes in this system. A more important difference, however, is that we probe the dynamic structure of the DW \textit{during} current injection rather than \textit{after} the current-induced displacement. Starting from a homogeneous magnetization state, we image the fastest DW front sweeping through the sample, which has opposite tilt with respect to the slowest DW front that survives in steady state conditions\cite{Ryu2012}. Accordingly, we find that the DW front in our measurements is orthogonal to the direction of largest DW velocity recently reported for Pt/Co/AlO$_x$\cite{Safeer2016}. These findings show that magnetization switching is boosted in the Pt/Co/AlO$_x$ dots by the favorable combination of the domain nucleation symmetry and DW propagation direction, which is such that the fastest DW front can sweep unhindered across the full extension of the dot. As an example of a less favourable case, we simulated the effect of a negative $B^{FL}$ (Fig. \ref{FL-simulations}a). Such a field would move the domain nucleation point to the opposite edge of the dot, while the direction of DW propagation along the $x$-axis would remain unaltered, leading to a much slower reversal dynamics.

\begin{figure}
	\centering
	\includegraphics[width=16cm]{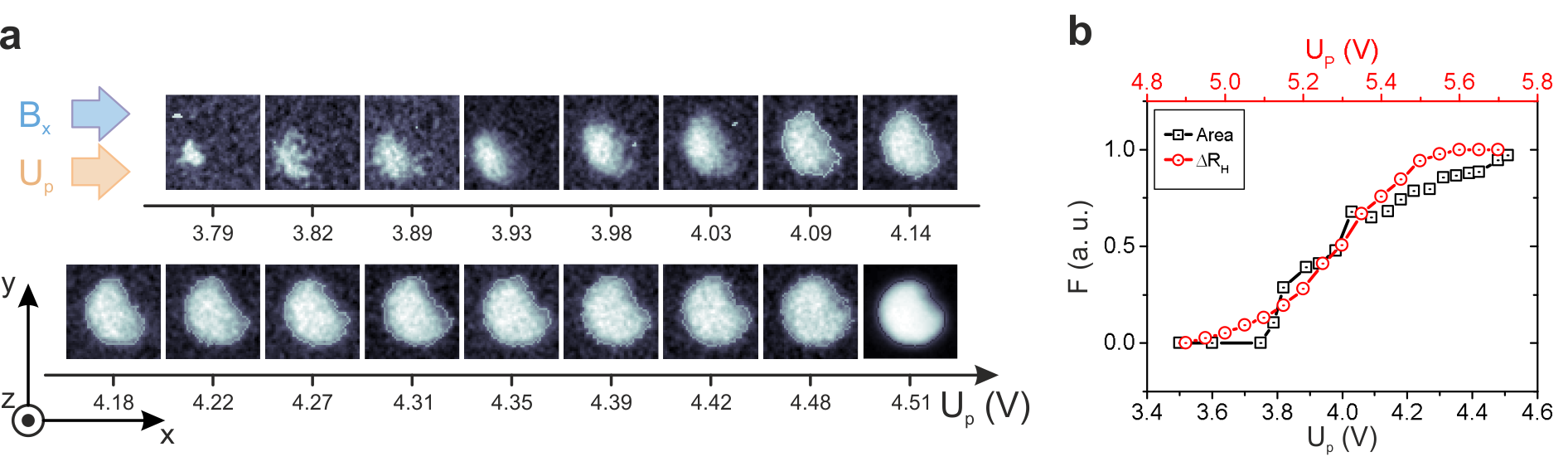}
	\caption{\textbf{Partial switching induced by sub-threshold current pulses.} \textbf{a,} Differential images showing the extent of magnetization reversal (in white) for pulses of increasing voltage amplitude. The differential contrast is obtained by averaging all the frames in a time-sequence after positive pulse injection and subtracting the average of all the frames after negative pulse injection. \textbf{b,} Comparison between the fractional reversed area estimated from the images in \textbf{a,} and the all-electrical switching measured by the Hall resistance on a replica dot. The Hall resistance data are averaged over 200 pulse cycles. The different voltage scale for the two measurements is due to the dispersion of the current in the branches of the Hall cross, which is absent in the STXM measurements.}
	\label{amp_dependence:pic}
\end{figure}
\paragraph*{Partial switching at sub-threshold current amplitude}
By calculating the time required for the DW front to cover the central region of the dots, we estimate that the DW velocity is of the order of 400 m/s, corresponding to about 100 m/s per $10^{8}$~A/cm$^2$ of injected current, in agreement with quasi-static measurements of DW displacements\cite{Miron2011b}. As magnetization reversal is deterministic and achieved by a single DW traversing the entire sample, the switching timescale is expected to be directly proportional to the lateral sample size. This can lead to switching times of less than 200~ps in structures that are smaller than 100~nm. Moreover, we find that pulses that are either shorter or weaker in amplitude compared to the threshold values required to achieve full switching consistently lead to the reversal of a fraction of the dot area. Figure~\ref{amp_dependence:pic}a shows the result of a series of switching measurements taken at increasing values of the voltage applied to the current line. Each frame is a differential image showing the average dot area that reversibly switches the magnetization upon applying positive and negative pulses. The reversed dot area increases monotonically with the pulse amplitude, as illustrated in Fig.~\ref{amp_dependence:pic}b, and correlates well with the remanent magnetization measured by the anomalous Hall effect on a replica dot. These measurements show that the critical switching current is mostly dependent on the DW mobility and sample dimensions rather than on the initial nucleation barrier. Further, our results agree with the absence of DW inertia reported in Pt/Co layers\cite{Vogel2012,Taniguchi2015}, consistently with the fast damping of magnetic excitations in SOT devices, and show that partial but reliable switching can be obtained also when working below the current amplitude required for full switching.

\paragraph*{Conclusions}
Controlling the speed of magnetic switching by electrical currents is one of the main challenges in nonvolatile memory technologies. Our results demonstrate reliable sub-ns magnetization reversal of perpendicularly magnetized Pt/Co dots induced by SOTs over more than $10^{12}$ switching cycles. Time-resolved STXM provides unprecedented insight into the spatial evolution of the magnetization and SOT-induced dynamics during the reversal process, revealing a four-fold asymmetry in the domain nucleation site at the dot edge and in the DW propagation direction, depending on the relative alignment of current and external field. The fast direction of DW motion is diagonal to the current, opposite to the one observed at steady state in racetrack structures. Our findings are complemented by pulsed switching Hall measurements and micromagnetic simulations, which disclose the role of the FL torque in determining the nucleation and DW dynamics in addition to the DL torque and DMI. We anticipate that tuning the amplitude and sign of the FL torque independently of the DL torque, or engineering structures with a significant Oersted field, may lead to improvements in the efficiency of SOT switching. Moreover, as the switching unfolds along a reproducible and deterministic path, the timing and the extent of magnetization reversal can be reliably controlled by the amplitude and duration of the current pulses as well as by the sample dimensions.


\bigskip \subsubsection*{Acknowledgments}
This work was funded by the Swiss National Science Foundation (Grant No. 200021-153404). We acknowledge fruitful discussions with J. St\"{o}hr and L. Buda-Prejbeanu.

\bigskip \subsubsection*{Author contributions}
M.B., K.G., and P.G. planned the experiments and analyzed the data. M.B., M.G., and J.F. fabricated the samples. M.B., K.G., J.M., C.O.A., E.G., C.M., C.S., Y.A., S.F., S.W., and J.R. implemented the current pump / x-ray probe scheme and performed the STXM experiments. M.B. carried out the electrical measurements and the micromagnetic simulations. M.B. and P.G. wrote the manuscript. All authors discussed the data and commented on the manuscript.

\bigskip \subsubsection*{Additional information}
Correspondence and requests for materials should be addressed to \\
M.B. (\verb"manuel.baumgartner@mat.ethz.ch") \\ and P.G. (\verb"pietro.gambardella@mat.ethz.ch").

\subsection*{Methods}
\paragraph*{Sample fabrication}
The samples were fabricated on 200~nm thick Si$_3$N$_4$ membranes supported by a rigid Si frame. The Pt(\unit[5]{nm})/Co(\unit[1]{nm})/Al(\unit[1.6]{nm}) layer was deposited by dc magnetron sputtering at a base pressure of $\unit[4\cdot10^{-8}]{mTorr}$ and the Al cap layer was subsequently oxidized in an O$_2$ atmosphere of \unit[13]{mTorr} for \unit[35]{s} and a power of \unit[36]{W}. Circular Co dots with 500~nm diameter and 750~nm wide Pt current lines were patterned by electron beam lithography followed by an ion milling process optimized to stop at the Pt/Co interface and at the Si$_{3}$N$_{4}$/Pt interface, respectively. Au contacts for current injection were fabricated in the proximity of the Co dots using UV-photo lithography and lift-off. Finally, a $\unit[100]{nm}$ thick Al layer was deposited by sputtering on the backside of the membrane in order to efficiently dissipate the heat originating from current-induced Joule heating.

\paragraph*{Scanning transmission x-ray microscopy}
The magnetization of the dots was imaged by time-resolved scanning transmission x-ray microscopy at the PolLux beamline of the Swiss Light Source\cite{Raabe2008} using a current pump / x-ray probe scheme. This method uses a Fresnel zone plate to focus a monochromatic x-ray beam onto the sample, which is scanned through the x-ray focus while an avalanche photodiode collects the transmitted photon intensity. Typical dot images are obtained by recording the x-ray intensity over a two-dimensional $64 \times 64$ pixel array with about \unit[25]{nm} lateral resolution. The pump current consists of a bipolar pulse sequence that has a total length of \unit[102.1]{ns}. Each unit sequence includes one positive and one negative pulse spaced by $\unit[50]{ns}$, the length and amplitude of which can be tuned independently from each other. The magnetization state is probed by exploiting the XMCD contrast at the Co $L_{3}$ absorption edge ($E = \unit[779]{eV}$) using left circularly polarized light at normal incidence. The probe x-ray beam has a \unit[70]{ps} wide bunch structure with a repetition rate of \unit[500]{MHz}. The magnetization is thus sampled every \unit[2]{ns} results in a time resolution of \unit[100]{ps} after the injection of 20 successive unit-sequences by routing the intensity of each probing event to a counting register. The current pulse sequences run continuously during the measurements, synchronized to the x-ray pulses, and are monitored by means of a \unit[-20]{dB} pick-off tee connected to an oscilloscope. The amplitude of the pulses is reported in voltage units, with $U_p = \unit[1]{V}$ corresponding to a current density $j_p = \unit[8.4\cdot10^7]{A/cm^2}$ in the Pt line.

\paragraph*{Electrical measurements}
Replicas of the dots were fabricated on Si$_3$N$_4$ membranes with the Pt underlayer patterned in the shape of a Hall cross (see Supplementary Fig. 1). These samples were deposited and processed together with those used for the x-ray measurements. We performed all-electrical pulsed switching measurements using the anomalous Hall resistance to detect the magnetization state after current injection\cite{Garello2014}.
The Hall measurements are averaged over 200 set pulses with length \unit[2]{ns}, each followed by a stronger reset pulse to ensure that the magnetization reverts to a homogeneous state before injection of the next set pulse. The Harmonic Hall voltage measurements of the SOTs\cite{Garello2013,Kim2013} were performed on $\unit[5]{\mu m}$ wide Pt/Co/AlO$_x$ Hall bars, which were also deposited and processed simultaneously with the dots. These measurements, corrected for a small contribution due to the Nernst effect\cite{Avci2014b}, give $B^{DL} = \unit[18]{mT}$ and $B^{FL} = B^{FL}_0 + B^{FL}_2 \sin^2 \theta$, where $\theta$ is the polar angle of the magnetization relative to the $z$-axis, $B^{FL}_0 = \unit[10.2]{mT}$, and $B^{FL}_2 = \unit[18]{mT}$ for a current density of $\unit[10^8]{A/cm^2}$ (see Supplementary Information).

\paragraph*{Micromagnetic simulations} Micromagnetic simulations based on the integration of the Landau-Lifshitz Gilbert equation were performed using the object oriented micromagnetic framework (OOMMF) code\cite{Donahue1999a}, including the DMI extension module\cite{Rohart2013a} as well as the DL and FL SOTs.
All simulations of the \unit[500]{nm} wide and \unit[1]{nm} thick Co dot were carried out using a cell size of $\unit[4]{nm} \times \unit[4]{nm}  \times \unit[1]{nm}$ with the following material parameters: saturation magnetization $M_s = \unit[900]{kA/m}$, exchange coupling constant $A_{ex} = \unit[10^{-11}]{A/m}$, uniaxial anisotropy energy $K_u = \unit[657]{kJ/m^3}$, DMI $D = \unit[1.2]{mJ/m^3}$, and damping constant $\alpha = 0.5$. The magnitude of the DL and FL torques corresponds to the values measured by the harmonic Hall voltage method; the in-plane bias field is set to $B_{x} = \unit[93]{mT}$. The simulated current pulse has an amplitude of $j_p = \unit[4.5\cdot10^8]{A/cm^2}$ and a duration of $\tau_p = \unit[2]{ns}$. All simulations were carried out at zero temperature.

\pagebreak

\setcounter{equation}{0}
\setcounter{figure}{0}
\setcounter{table}{0}
\setcounter{page}{1}
\makeatletter
\renewcommand{\thesection}{SI\arabic{section}}
\renewcommand{\theequation}{S\arabic{equation}}
\renewcommand{\thefigure}{S\arabic{figure}}
$\renewcommand{\bibnumfmt}[1]{[S#1]}
$\renewcommand{\citenumfont}[1]{S#1}

\begin{center}
	
	\textbf{\large{SUPPLEMENTARY INFORMATION\\}}
	\bigskip
	\textbf{	
		Time- and spatially-resolved magnetization dynamics driven by spin-orbit torques\\
	}
	\bigskip
	\normalsize{
	M. Baumgartner,$^{1}$ K. Garello,$^{1,2}$ J. Mendil,$^{1}$ C. O. Avci,$^{1}$ E. Grimaldi,$^{1}$\\ C. Murer,$^{1}$ J. Feng,$^{1}$ M. Gabureac,$^{1}$ C. Stamm,$^{1}$ Y. Acremann,$^{3}$\\S. Finizio,$^{4}$ S. Wintz,$^{4}$ J. Raabe,$^{4}$ P. Gambardella$^{1}$\\
	\bigskip
	\small{$^{1}$Department of Materials, ETH Z\"urich, 8093 Z\"urich, Switzerland,}\\
	\small{$^{2}$IMEC, Kapeldreef 75, 3001 Leuven, Belgium,}\\
	\small{$^{3}$Laboratory for Solid State Physics, ETH Z\"urich, 8093 Z\"urich, Switzerland,}\\
	\small{$^{4}$Paul Scherrer Institute, 5232 Villigen PSI, Switzerland\\}}
		
\end{center}

	\bigskip
	\bigskip
	\tableofcontents{}
	\bigskip
	\bigskip
	\clearpage

	\section{Magnetic characterization and all-electrical pulsed switching measurements}\label{pulsed}
\begin{figure}[b!]
		\centering
		\includegraphics[width=16cm]{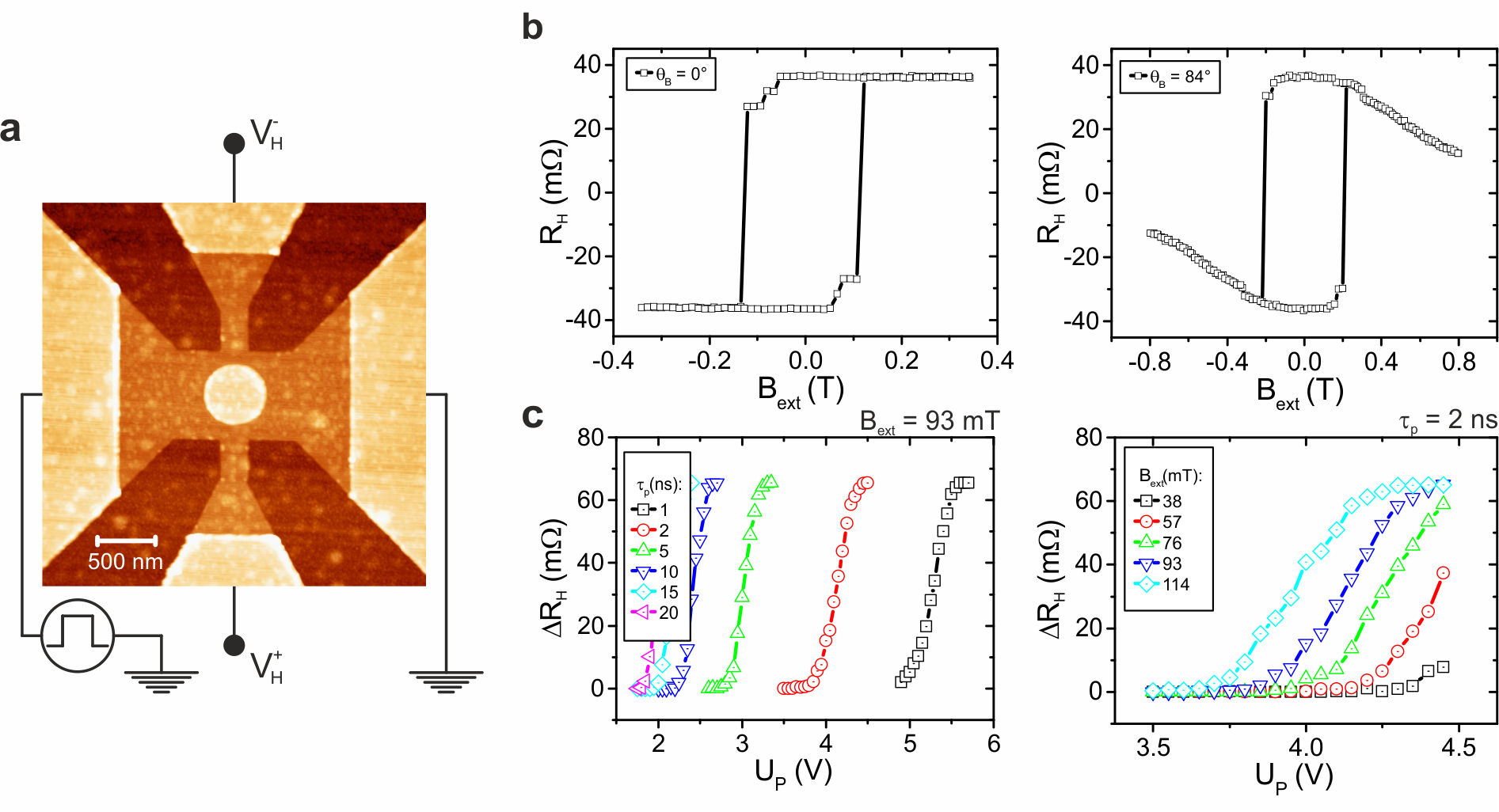}
		\caption{\textbf{Magnetic characterization and all-electrical pulsed switching measurements.} \textbf{a,} Atomic force microscopy image showing a \unit[500]{nm} Co dot on a \unit[750]{nm} wide Pt Hall cross grown on a Si$_3$N$_4$ membrane. \textbf{b,} $R_{H}$ measured as a function of external field applied out-of-plane (left panel, $\theta_B =0^{\circ}$) and nearly in-plane (right panel, $\theta_B =84^{\circ}$). \textbf{c,} Pulsed magnetization switching measurements as a function of pulse length (left panel) and in-plane bias field (right panel). In both cases the bias field is applied at an angle $\theta_B =89^{\circ}$ along the current line.}\label{transportCharacterization:pic}
	\end{figure}

	Figure~\ref{transportCharacterization:pic}a shows an atomic force microscope image of a 500~nm wide Co dot on top of a Pt Hall cross deposited on a 200~nm thick Si$_3$N$_4$ membrane. The Co dot is 1~nm thick, capped by 2~nm of oxidized Al, and the Pt line is 5~nm thick. This type of samples are replicas of those employed for the STXM measurements, which we used to characterize the magnetic properties and to perform all-electrical pulsed switching experiments by probing the anomalous Hall effect\cite{Garello2014}. Figure~\ref{transportCharacterization:pic}b shows the Hall resistance $R_{H}=R_{AHE}\cos\theta = R_{AHE} M_z$ measured as a function of the external field $B_{ext}$ applied out-of-plane ($\theta_B =0^{\circ}$) and nearly in-plane ($\theta_B =84^{\circ}$). Here, $\theta$ and $\theta_B$ are the polar angles of the magnetization and applied field, respectively, and $R_{AHE}$ is the anomalous Hall coefficient. The square loop measured at $\theta_B =0^{\circ}$ shows that the dots have strong perpendicular magnetic anisotropy, with an effective anisotropy field given by $B_K = B_{ext} (\sin\theta_{B}/\sin\theta -\cos\theta_{B}/\cos\theta) \approx \unit[660]{mT}$.
	
	Figure~\ref{transportCharacterization:pic}c shows pulsed switching measurements as a function of the pulse voltage amplitude $U_p$ (\unit[1]{V} corresponds to $j_p = \unit[8.4 \cdot 10^8]{A/cm^2}$ in the Pt line) for different pulse widths $\tau_p$ and external fields applied at an angle $\theta_B =89^{\circ}$, almost parallel to the current line. Each point represents the difference of the Hall resistance measured after and before the pulse averaged over 200 switching attempts starting from a fully saturated magnetization state. We observe reliable switching at pulse voltages and widths comparable to those used for the STXM measurements, the main difference being that the $U_p$ values in the Hall measurements are about $\unit[20]{\%}$ larger compared to the x-ray measurements due to the dispersion of the current in the Hall cross. In agreement with previous studies\cite{Garello2014}, we find that the threshold pulse amplitude to achieve switching decreases with increasing pulse width and bias field.

	\section{Harmonic Hall voltage measurements of the spin-orbit torques}\label{HHV}
	\begin{wrapfigure}{r}{0.35\textwidth}
		\centering
		\includegraphics[width=5cm]{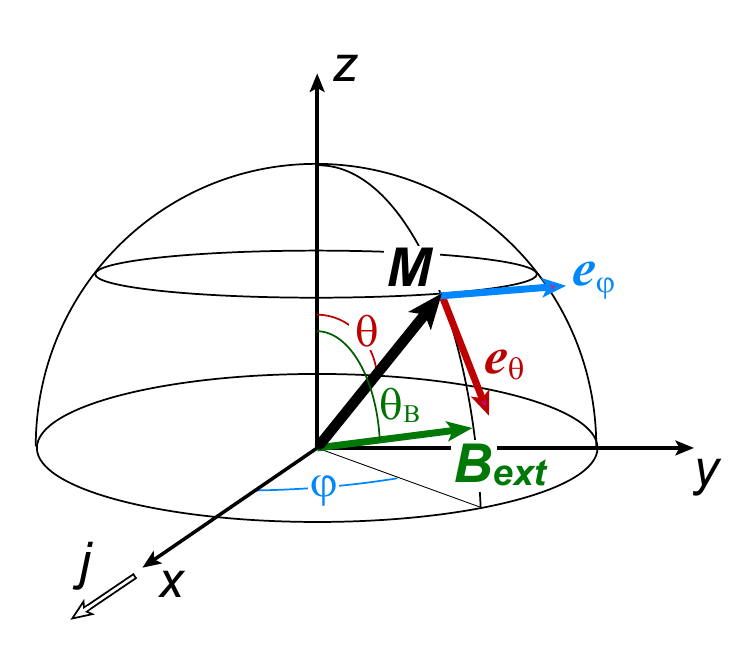}
		\caption{Coordinate scheme.}
		\label{Coordinates:pic}
	\end{wrapfigure}
\begin{figure}[b!]
		\centering
		\includegraphics[width=16cm]{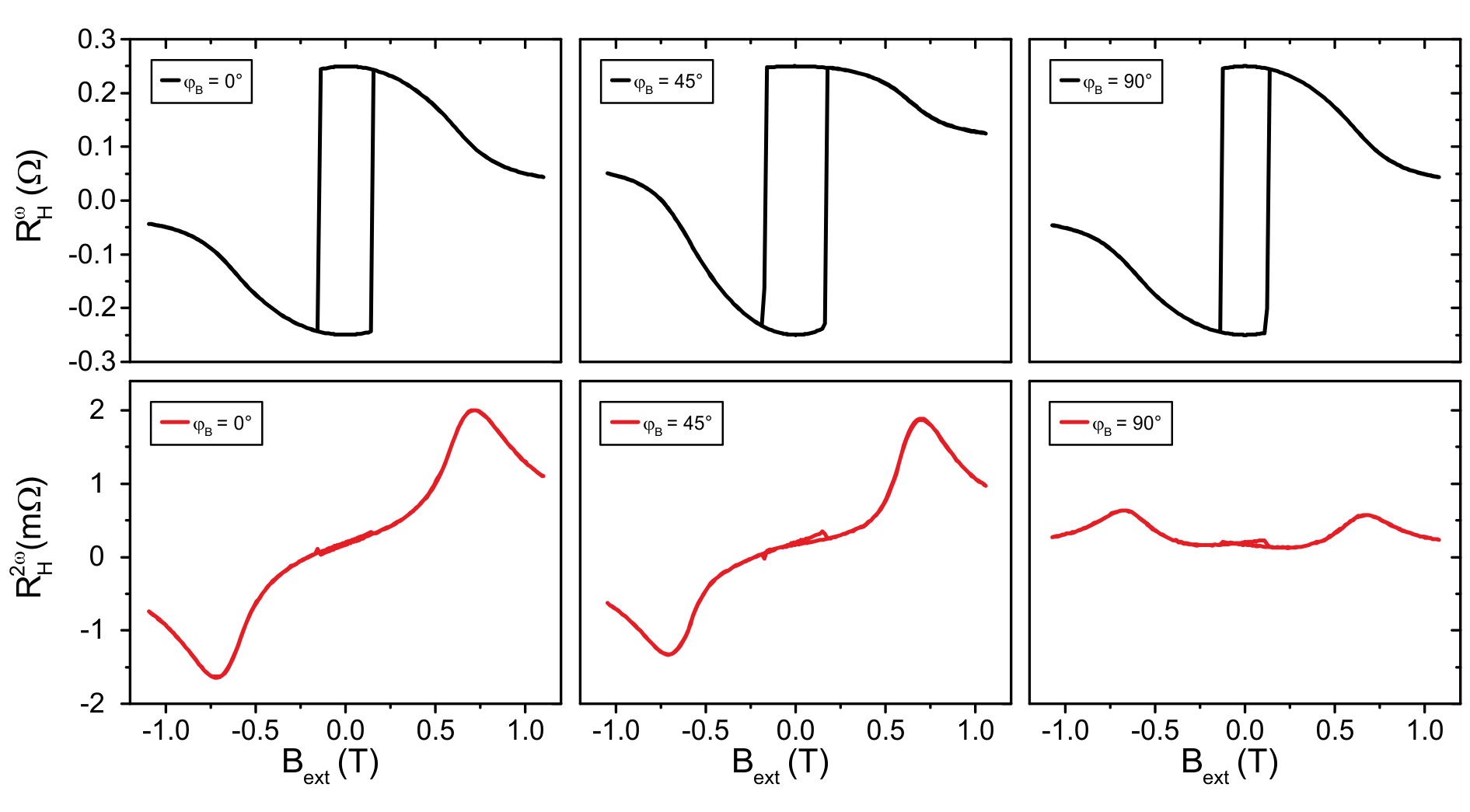}
		\vspace{-0.4 cm}
		\caption{\textbf{Harmonic Hall voltage measurements of the SOTs.} Upper row: First harmonic Hall resistance signals obtained for $\varphi_B = \unit[0]{^\circ}$, $\varphi_B = \unit[45]{^\circ}$ and $\varphi_B = \unit[90]{^\circ}$ at constant $\theta_B = \unit[86]{^\circ}$. Lower row: Corresponding second harmonic signal.}
		\label{SOTcharacterization:pic}
	\end{figure}

	In order to determine the SOTs and the effective perpendicular anisotropy field in our structures, we performed adiabatic harmonic Hall voltage measurements\cite{Garello2013, Kim2013} using an ac current of frequency $\omega =2\pi\cdot 10$~Hz and a current density of $j = \unit[10^7]{A/cm^2}$. The sample is a $\unit[5]{\mu m}$ wide Hall bar which has been deposited on Si$_3$N$_4$ at the same time as the Co dots used for the time-resolved switching experiments described in the main text.
	The injection of the ac current into the Pt/Co bilayers generates a Hall voltage perpendicular to the current direction, which depends on the orientation of the magnetization due to the anomalous and planar Hall effects. The Hall resistance $R_H = V_H/I = R_H^\omega + R_H^{2\omega}$ includes a first and a second harmonic component. The first harmonic component, shown in the top panels of Fig.~\ref{SOTcharacterization:pic}, is equivalent to the dc Hall resistance, which depends on the equilibrium position of the magnetization as
	\begin{equation}
	R_{H}^\omega= R_{AHE}\cos\theta + R_{PHE}\sin^2\theta\sin(2\varphi),
	\label{firstHarmonic:eq}
	\end{equation}
	where $R_{PHE}$ is the planar Hall coefficient and $\varphi$ the azimuthal angle describing the in-plane orientation of the magnetization relative to the $x$-axis (see Fig.~\ref{Coordinates:pic}). The second harmonic signal, which is proportional to the current, depends on the SOT-induced reorientation of the magnetization about the equilibrium position as well as on magnetothermal effects scaling with $I^2$. The general form of $R_{H}^{2\omega}$ is given by\cite{Garello2013, Avci2014b}
	\begin{equation}
	R_{H}^{2\omega}=\left[ R_{AHE} - 2R_{PHE}\cos\theta\sin(2\varphi)\right] \dfrac{d\cos\theta}{dB_{ext}}\dfrac{B_\theta^I}{\sin(\theta_B-\theta)} + R_{PHE}\sin^2\theta\dfrac{2\cos(2\varphi)}{B_{ext}\sin\theta_B}B_\varphi^I + R_{\nabla T}^{2\omega},
	\label{secondHarmonic:eq}
	\end{equation}
	where $B_\theta^I$ and $B_\varphi^I$ are the polar and azimuthal components of the total current-induced field $\mathbf{B}^I = \mathbf{B}^{DL}+\mathbf{B}^{FL}+\mathbf{B}^{Oe}$. By performing three independent measurements of $R_{H}^{\omega}$ and $R_{H}^{2\omega}$ at different angles (for instance at $\varphi_B = \unit[0]{^\circ}$, $\unit[45]{^\circ}$, $\unit[90]{^\circ}$, as shown in Fig.~\ref{SOTcharacterization:pic}) with a fixed $\theta_B$, these components can be directly linked to the SOTs expressed in spherical coordinates:
	\begin{equation}
	\mathbf{B}^{FL} = 		B_\theta^{FL}\cos\theta\sin\varphi\mathbf{e}_\theta+B_\varphi^{FL}\cos{\varphi}\mathbf{e}_\varphi,
	\label{B_FL:eq}
	\end{equation}	
	\begin{equation}
	\mathbf{B}^{DL} = B_\theta^{DL}\cos\varphi\mathbf{e}_\theta-B_\varphi^{DL}\cos\theta\sin\varphi\mathbf{e}_\varphi,
	\label{B_DL:eq}
	\end{equation}
	where $\mathbf{e}_{\varphi}$ and $\mathbf{e}_{\theta}$ are the azimuthal and polar unit vectors. The last term in Eq.~\eqref{secondHarmonic:eq} is the magnetothermal Hall resistance, which is mainly due to the anomalous Nernst effect induced by Joule heating. This term must be taken into account for accurate SOT quantification. The second harmonic components due to $R_{\nabla T}^{2\omega}$ and the current-induced fields are separated by analyzing the dependence of $R_{H}^{2\omega}$ on $B_{ext}$during a field scan with $\theta_B=\unit[90]{^\circ}$ and $\varphi_B=\unit[45]{^\circ}$. As the magnetization saturates in the sample plane at high fields, the susceptibility to the SOT decreases and $R_H^{2\omega} \rightarrow R_{\nabla T}^{2\omega}$. Plotting $R_H^{2\omega}$ as a function of $1/(B_{ext}-B_K)$ allows us to deduce $R_{\nabla T}^{2\omega}$ from the intercept of the coordinate axis\cite{Avci2014b}. In the present case, accounting for $R_{\nabla T}^{2\omega}$ leads to a reduction of the SOTs of about 20 \% (15 \%) for the FL (DL) torque relative to the values derived from the as-measured $R_{H}^{2\omega}$. Following this procedure we find that $B^{DL} = \unit[18]{mT}$, and $B^{FL} = B^{FL,Oe}_0 + B^{FL}_2 \sin^2 \theta$, where $B^{FL,Oe}_0 = \unit[10.2]{mT}$ and $B^{FL}_2 = \unit[18]{mT}$ for a current density of $\unit[10^8]{A/cm^2}$. The Oersted field contribution is evaluated in Sect.~\ref{Oersted}.

	\section{Dependence of the switching speed on the external field $B_x$}\label{Bx}
	\begin{figure}[b!]
		\centering
		\includegraphics[width=16cm]{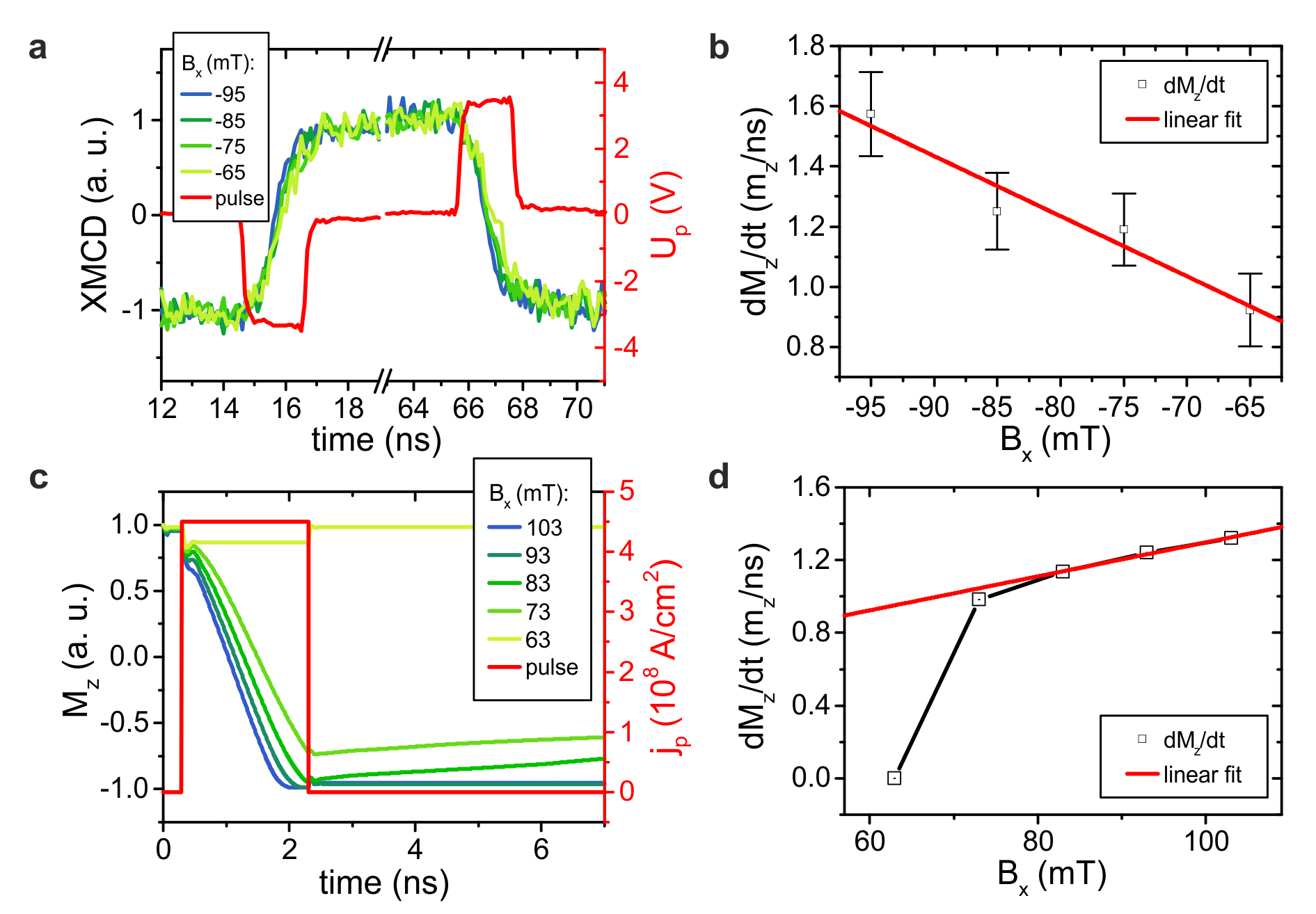}
		\caption{\textbf{Effect of the external in-plane field on the switching speed.} \textbf{a,} Measured XMCD time traces as a function of in-plane field $B_x$. \textbf{b,} Average slope of the rising and falling edges during the reversal plotted as function of $B_x$. \textbf{c,} Simulated time traces of $M_z$ as a function of $B_x$. The slow dynamics after the current pulse is due the DW reorientation after incomplete magnetization reversal. \textbf{d,} Reversal rate $\frac{dM_z}{dt}$ evaluated for the traces shown in \textbf{c} during the current pulse.}
		\label{fieldDependence:pic}
	\end{figure}
	Figure \ref{fieldDependence:pic}a shows the XMCD time traces of a Co dot during the injection of $\unit[2]{ns}$ long current pulses of amplitudes $U_p=$\unit[-3.35]{V} and \unit[3.42]{V} at different values of the applied field $B_x$. In the following, we will limit the discussion to negative fields but emphasize that the same arguments are consistent with positive bias fields. We find that $B_x$ has an influence on the reversal speed, namely that switching occurs faster at high field values. This is shown in Fig. \ref{fieldDependence:pic}b where the average slope of the XMCD time traces during the reversal is plotted against $B_x$. With increasing $B_x$, $M_z$ decreases while $|M_x|$ increases on one side of the dot, which favours magnetization reversal.
	
	Our findings are supported by micromagnetic simulations (see Methods Section) in which we analyze the influence of $B_x$ on the reversal speed. Figure \ref{fieldDependence:pic}c shows the simulated time traces of $M_z$ following the injection of a current pulse of amplitude $j_p = \unit[4.5\cdot10^8]{A/cm^2}$ for different values of $B_x$.
	The simulations show that there is a threshold bias field $63~\mathrm{mT } < B_x < 73$~mT below which reversal does not take place.
	Above this field, the DW velocity increases with $B_x$ as the magnetization tilts more towards the sample plane and the energy barrier for DW propagation reduces, in agreement with the experiment. We further note that the experimental threshold field can be significantly lower compared to the simulations owing to Joule heating, which assists the nucleation of reversed domains.

	\section{Macrospin model of magnetization reversal at different edge positions}\label{macrospin}
	\begin{figure}[b!]
		\centering
		\includegraphics[width=16cm]{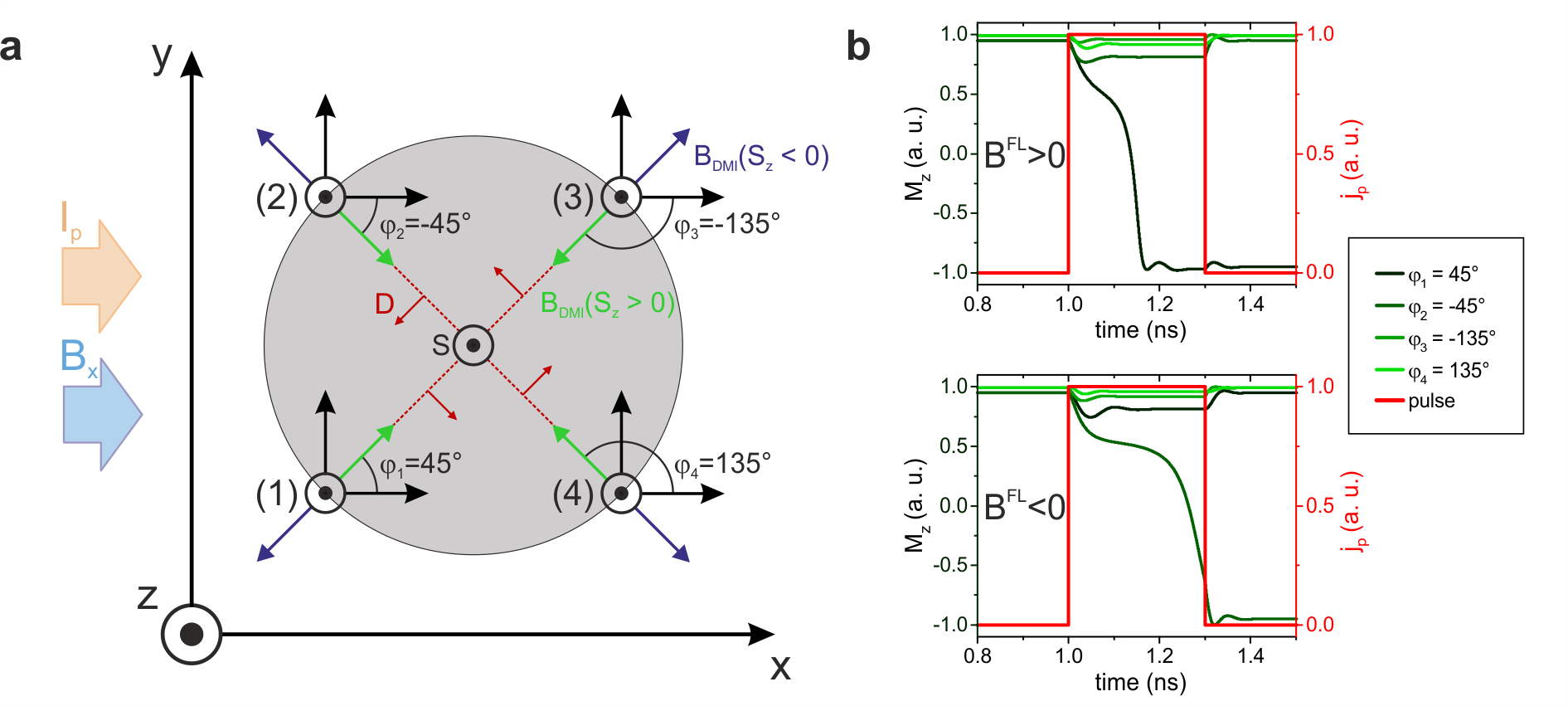}
		\caption{\textbf{Macrospin model of edge magnetization reversal.} \textbf{a,} Schematic representation of four non-interacting spins located at distinct positions around the dot. The DMI is taken into account as an effective in-plane field, the sign of which depends on the orientation of the fictitious central spin. \textbf{b,} $z$-component of the four macrospins during the application of a current pulse for $B^{FL}>0$ (upper panel, corresponding to the experimental situation) and $B^{FL}<0$ (lower panel).}
		\label{DetailedDW_nucleation:pic}
	\end{figure}
	We provide a simplified and intuitive model of the edge nucleation by simulating the reversal process of four non-interacting single spins located at positions (1) - (4) around the Co dot, as illustrated in Fig.~\ref{DetailedDW_nucleation:pic}a. The DMI is included as an effective field by considering the DMI-energy
	\begin{equation}
	E_{DMI}= \mathbf{D}\cdot\left(\mathbf{M}_i\times\mathbf{S} \right),
	\label{E-DMI:eq}
	\end{equation}
	where $\mathbf{M}_i$ is the magnetization vector of the spin at position ($i$), $\mathbf{S}$ represents a fictitious central spin of fixed orientation and $\mathbf{D}$ is the DMI-coupling vector. $\mathbf{D}$ lies in the sample plane and is perpendicular to the vector connecting the spins $\mathbf{M}_i$ and $\mathbf{S}$. It can thus be parametrized by an angle $\varphi_i$ as $\mathbf{D}=D(\sin\varphi_i, -\cos\varphi_i, 0)$. We then make use of the general relation $\mathbf{B}_{DMI}=-\nabla_\mathbf{M} E_{DMI}$ to express the DMI-energy as an effective field
	\begin{equation}
	\mathbf{B}_{DMI,i} = \text{sgn}\left(S_z\right)  2D
	\begin{pmatrix}
	\cos\varphi_i \\
	\sin\varphi_i \\
	0
	\end{pmatrix}.
	\end{equation}
	
	Finally, we perform a time-integral of the Landau-Lifshitz Gilbert equation to separately model the evolution of the magnetization at points (1) - (4) under the action of $B_{DMI,i}$, $B_{x}=\unit[100]{mT}$, effective anisotropy field $B_K = \unit[657]{mT}$, and the DL and FL torques measured in Section~\ref{HHV}. We choose the DMI-coupling strength $D$ such that it induces a spin
	
	
	canting at the sample edge similar to that reported in Ref.~\cite{Mikuszeit2015}. The canting angle at positions (1) and (2) under a positive bias field of $B_x = \unit[100]{mT}$ is $\unit[\approx18]{^\circ}$ for $D=\unit[60]{mT}$, whereas it is $\unit[\approx8]{^\circ}$ at positions (3) and (4). For brevity, we limit the discussion to $B_x > 0$ and positive current pulses, as indicated in Fig. \ref{DetailedDW_nucleation:pic}a.
	
	We consider first the case $S_z > 0$, which represents the dot in the up-state. For a given current, the simulated time traces of the magnetization at positions (1) - (4), reported in Fig.~\ref{DetailedDW_nucleation:pic}b, show that only the spin located at position (1) reverses, whereas, by inverting the sign of $B^{FL}$ and keeping $B^{DL}$ constant, only the spin at position (2) reverses. This behavior is in agreement with the nucleation point observed experimentally (Fig. 3b (II)). Inverting the sign of the central spin, such that $S_z < 0$, changes the direction of the effective DMI-field and makes position (3) equivalent to position (1) in the former case ($S_z > 0$). The effect of the FL torque at different positions can be simply rationalized by considering the $\theta$-component of the field $B^{FL}$:
	\begin{equation}
	{B}^{FL}_{\theta} =	\mathbf{B}^{FL}\cdot\mathbf{e}_\theta =
	\begin{pmatrix}
	0 \\
	B^{FL} \\
	0
	\end{pmatrix}
	\cdot
	\begin{pmatrix}
	\cos\theta\cos\varphi_i \\
	\cos\theta\sin\varphi_i \\
	-\sin\theta
	\end{pmatrix}= B^{FL}\cos\theta\sin\varphi_i,
	\label{B_FL-theta:eqn}
	\end{equation}
	where we see that ${B}^{FL}_{\theta}$ depends on $\varphi_i$ and points towards the $xy$-plane at position (1) and in the opposite direction at position (2), thus favoring the reversal of spin (1) relative to (2).

	\section{Effect of the Oersted field}\label{Oersted}
	\begin{figure}[b!]
		\centering
		\includegraphics[width=16cm]{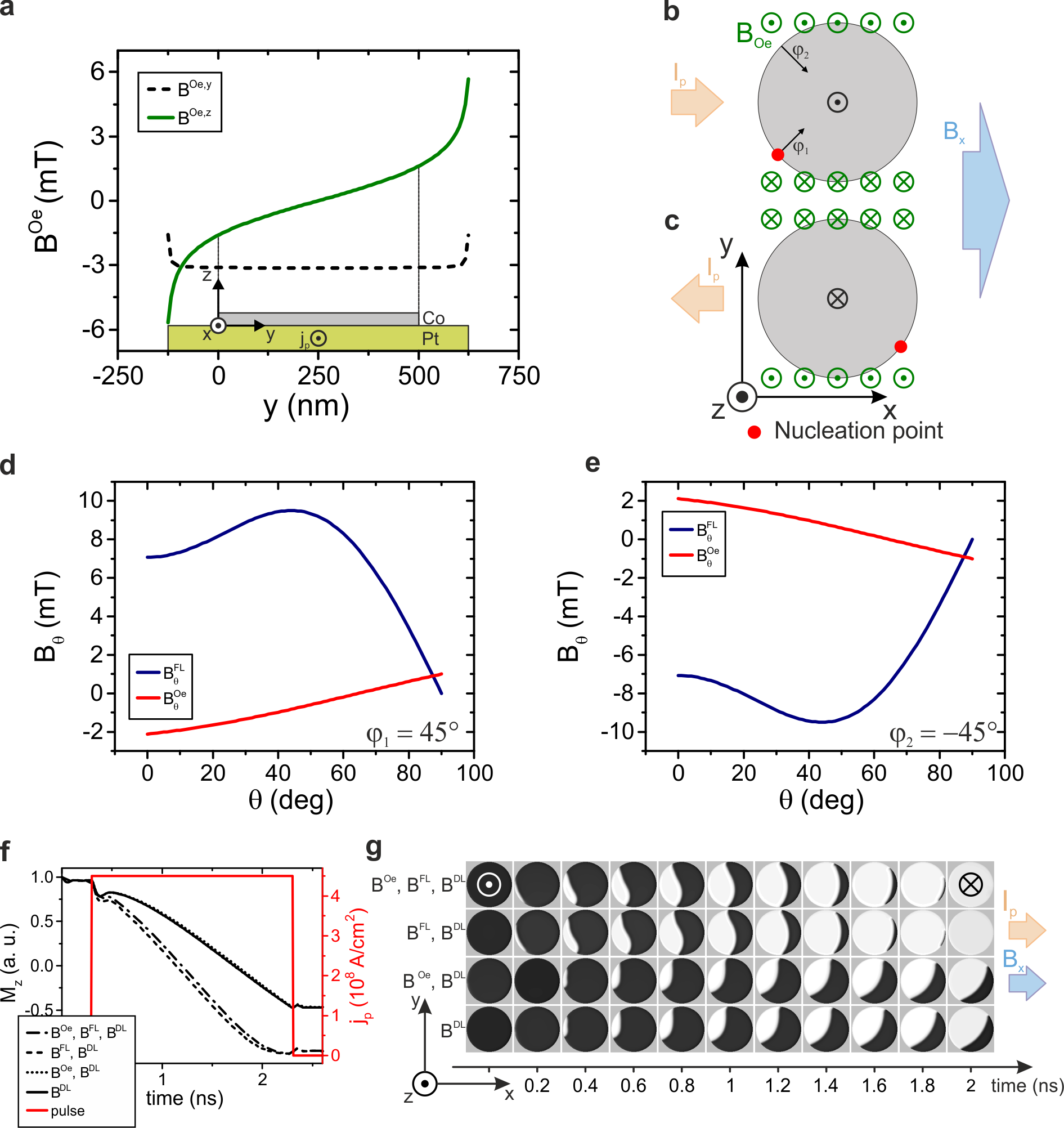}
		\caption{\textbf{Influence of the Oersted field on magnetization reversal.} \textbf{a,} Calculated Oersted field according to Eq.~\eqref{B_Oe:eqn} at a height of \unit[0.5]{nm} above the Pt line for a positive current pulse of amplitude $\unit[1\cdot10^8]{A/cm^2}$. \textbf{b,} Illustration of the nucleation process supported by the $z$-component of the Oersted field at positive and \textbf{c,} negative current. \textbf{d,} $\theta$-component of $B^{FL}$ and $B^{Oe}$ as a function of $\theta$ during a positive current pulse of amplitude $\unit[1\cdot10^8]{A/cm^2}$ for a spin located at the position (1) and \textbf{e,} (2) of the dot (see diagram in Fig.~\ref{DetailedDW_nucleation:pic}). \textbf{f,} Simulated time traces for different combinations of $B^{Oe}$, $B^{FL}$ and $B^{DL}$. \textbf{g} Corresponding snapshots during the magnetization reversal.}
		\label{OerstedField:pic}
	\end{figure}
	The current flowing in the Pt line produces an Oersted field with $y$- and $z$-components given by\cite{Hayashi2014}
	
	\begin{equation}
	B^{Oe}_y = -\dfrac{I}{wd}\left(U_y^+-D_y^++U_y^--D_y^-\right)\cdot 10 ^{-3} , \qquad B^{Oe}_z = -\dfrac{I}{wd}\left(U_z^+-D_z^+-U_z^-+D_z^-\right)\cdot 10 ^{-3},
	\label{B_Oe:eqn}
	\end{equation}
	where $I$ is the current, $w$ is the width of the Pt line and $d$ its thickness, respectively. The coefficients are defined as follows:
	\begin{equation}
	U_y^\pm = 2 (z +d) \arctan\left[\dfrac{w/2 \pm y}{z+d}\right] +(w/2\pm y) \ln\left[\left(w/2 \pm y \right)^2+(z+d)^2 \right],
	\end{equation}
	\begin{equation}
	D_y^\pm = 2 z \arctan\left[\dfrac{w/2 \pm y}{z}\right] +(w/2\pm y) \ln\left[\left(w/2 \pm y \right)^2+z^2 \right],
	\end{equation}
	\begin{equation}
	U_z^\pm = 2 (w/2 \pm y) \arctan\left[ \dfrac{z+d}{w/2 \pm y}\right] + (z+d)\ln\left[\left(w/2 \pm y \right)^2+(z+d)^2  \right],
	\end{equation}
	\begin{equation}
	D_z^\pm = 2 (w/2 \pm y) \arctan\left[ \dfrac{z}{w/2 \pm y}\right] + z\ln\left[\left(w/2 \pm y \right)^2+z^2  \right],
	\label{B_Oe_Coeff:eqn}
	\end{equation}
	with $z$ being the height above the surface of the Pt line.
	
	Figure~\ref{OerstedField:pic}a shows the Oersted field calculated according to Eq.~\ref{B_Oe:eqn}-\ref{B_Oe_Coeff:eqn} for a current density of $\unit[1\cdot10^8]{A/cm^2}$ flowing through the Pt layer, at $z = 0.5$~nm above the Pt surface. The $y$-component $B^{Oe}_y=-3$~mT is approximately constant over the dot surface and directed against $B^{FL}$. The largest $z$-component is found at the two extrema of the dots closest to the edge of the Pt line, where $|B^{Oe}_z| = \unit[1.6]{mT}$. The sign of $B^{Oe}_z$ is opposite on opposite sides of the dot, such that this field can also induce a top/bottom edge asymmetry, supporting the nucleation process induced by $B^{FL}$, as shown in Figs.~\ref{OerstedField:pic}b and c.
	
	To analyze the role played by the Oersted field and FL torque in the nucleation, we plot in Fig.~\ref{OerstedField:pic}d the $\theta$-components of $B^{FL}$ and $B^{Oe}$ as a function of $\theta$. The fields are evaluated at position (1) in Fig.~\ref{DetailedDW_nucleation:pic}, where the magnetization is tilted by the DMI and external field along $\varphi = 45^{\circ}$. At this position, the total $B_{\theta}$ is positive for all $\theta$, parallel to $\mathbf{e}_{\theta}$ (Fig.~\ref{Coordinates:pic}), and thus favours the rotation of the magnetization from up to down. However, without $B^{FL}$ magnetization reversal would not be favoured at this position since $B^{Oe}_\theta$ is negative over a large range of $\theta$. The opposite argumentation is valid for the magnetization at position (2) in Fig.~\ref{DetailedDW_nucleation:pic}a, where the reversal is favoured by $B^{Oe}$ and hindered by $B^{FL}$ (Fig. \ref{OerstedField:pic}e). As (1) corresponds to the experimentally observed nucleation, we conclude that $B^{FL}$ is the main cause of the nucleation asymmetry. It is important to note that $B^{Oe}_y$ is approximately constant over the dot area whereas $B^{Oe}_z$ increases rapidly near the Pt line edge.

	This implies that magnetization nucleation will be eventually also assisted by the Oersted field for Co dots with a diameter comparable to the Pt line width. To demonstrate the effect of $B^{Oe}$ on the switching process, we performed a series of micromagnetic simulations with
	$i$) $B^{Oe}\neq 0$ and $B^{FL} \neq 0$, $ii$) $B^{Oe}= 0$ and $B^{FL} \neq 0$, $iii$) $B^{Oe}\neq 0$ and $B^{FL}=0$ and $iv$) $B^{Oe}= 0$ and $B^{FL}=0$, the results of which are reported in Figs.~\ref{OerstedField:pic}f and g. Without $B^{FL}$ the switching efficiency is strongly reduced, whereas the addition of $B^{Oe}$ to $B^{FL}$ has only a small influence on the reversal mechanism for the given device geometry. The simulations also show how the nucleation point moves up along the edge for $B^{FL}=0$.

	\section{Switching of defective dots}\label{defect}
\begin{figure}[t!]
		\centering
		\includegraphics[width=16cm]{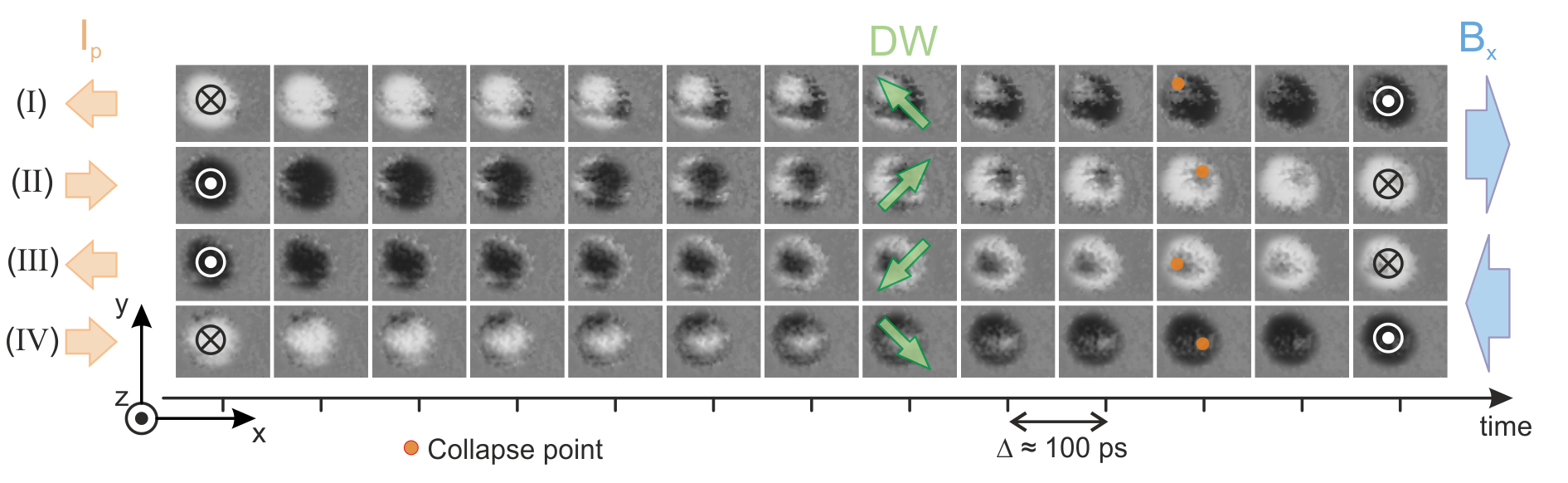}
		\caption{\textbf{Switching in the presence of defects.} Images taken at intervals of 100~ps during the injection of 2~ns long current pulses. DW nucleation is observed at different sites. The point at which the reversed domain front collapses (orange dot) is off-centred, in agreement with the preferential DW propagation direction discussed in the main text (green arrows).}
		\label{defectSwitching:pic}
	\end{figure}

	All of the measured dots show reproducible switching behavior rather than thermally-induced stochastic reversal, which would result in uniform STXM contrast during the reversal process. However, approximately 50\% of the dots showed defect-promoted switching, likely due to inhomogeneities occurred during the patterning process. Such samples are distinguished by the ones presented in the main text by the fact that nucleation always starts from a given point or region rather than alternating between the four dot quadrants. Figure~\ref{defectSwitching:pic} shows a case of defect-promoted switching in which domain nucleation starts at several points around the Co dot edge. Due to the stroboscopic nature of our measurements, we cannot tell whether the nucleation starts simultaneously at different locations or is randomly distributed over the dot edge. However, we do observe that once a domain has nucleated, the favoured DW propagation direction remains the same (green arrows in Figure~\ref{defectSwitching:pic}) and is comparable to the switching behaviour reported in Fig. 3 of the main text. The point at which the reversed domain front collapses (orange dot) is therefore off-centred. Importantly, we find that in all cases the reversal process is robust with respect to the presence of defects, which makes SOT-induced magnetization switching very versatile for applications.

	\section{Comparison between filtered and raw data}
\begin{figure}[h!]
		\centering
		\includegraphics[width=16cm]{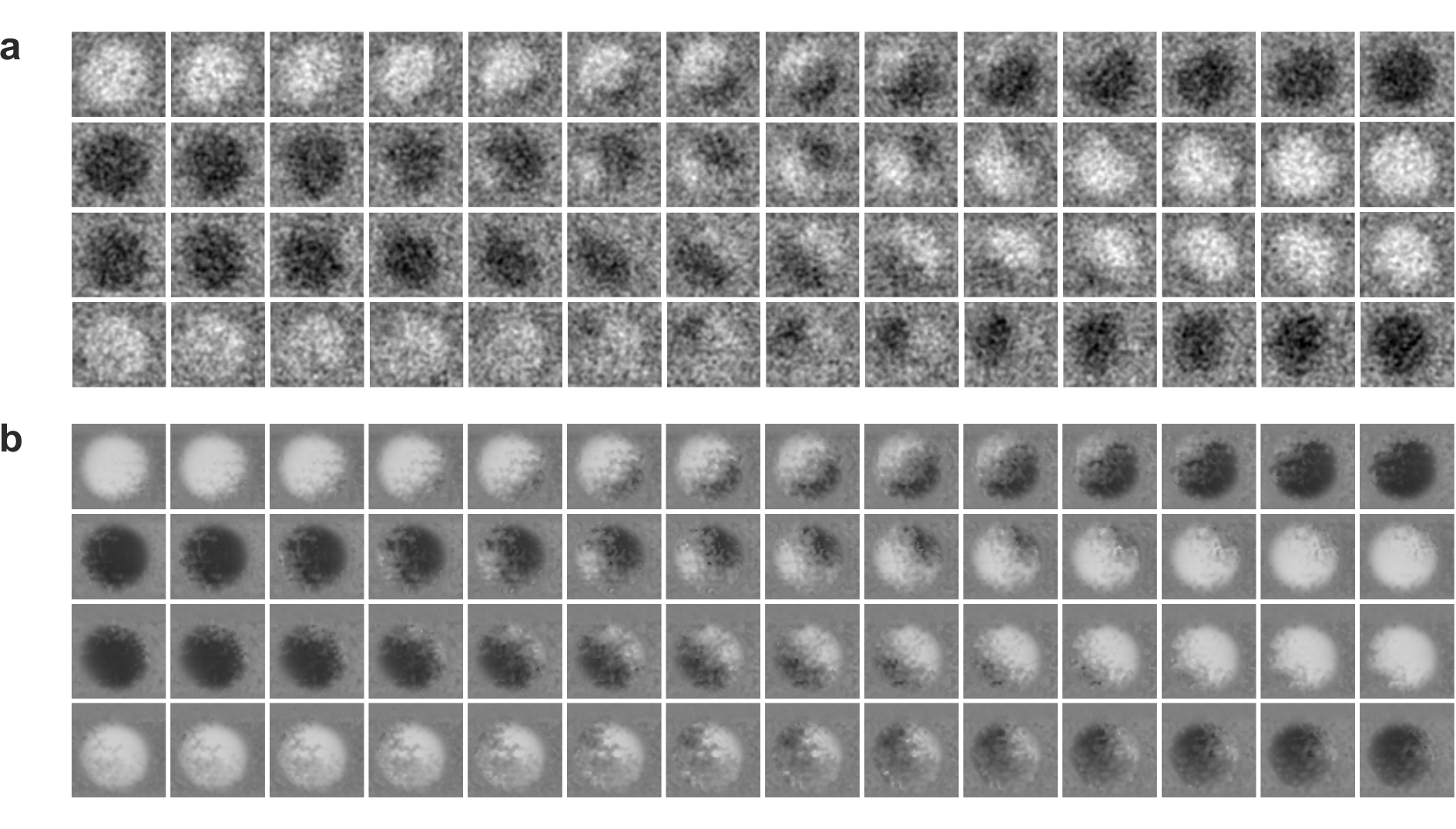}
		\caption{\textbf{Comparison of the raw and filtered data shown in Fig. 3.} \textbf{a,} Raw and \textbf{b,} low-pass filtered images.}
		\label{RAWvsFilteredDMI:pic}
	\end{figure}

\begin{figure}[h!]
		\centering
		\includegraphics[width=16cm]{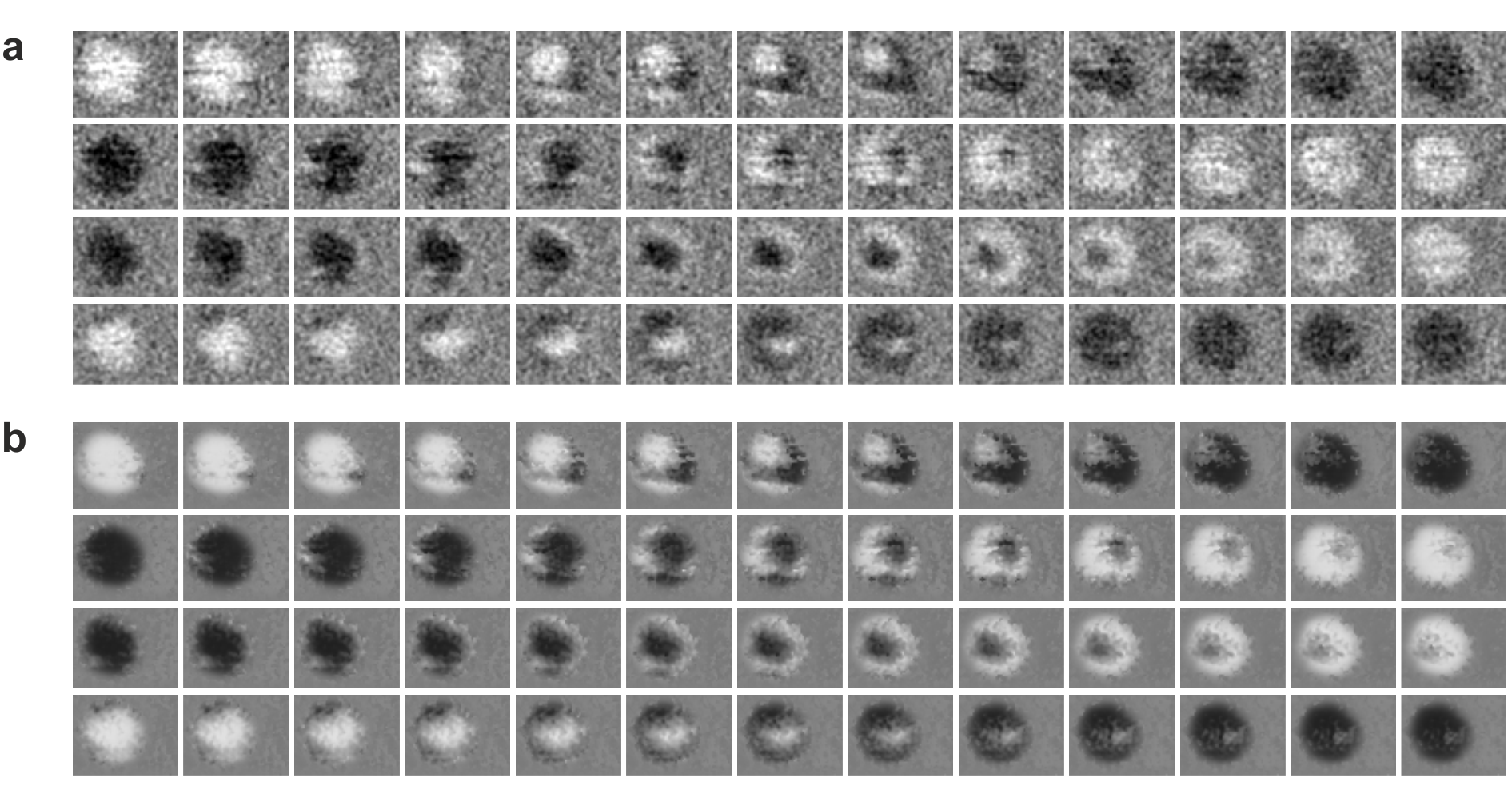}
		\caption{\textbf{Comparison of the raw and filtered data shown in Fig.~\ref{defectSwitching:pic}.} \textbf{a,} Raw and \textbf{b,} low-pass filtered images.}
		\label{RAWvsFilteredDotCollapse:pic}
	\end{figure}

	Given the very low photon count rate at the detector position and the relatively low x-ray absorption contrast corresponding to the 1~nm thick Co layer, the acquisition time of a single sequence of images with 100~ps resolution requires several hours of integration. In order to enhance the magnetic contrast, the intensity of each frame in Fig. 3 of the main text and Fig.~\ref{defectSwitching:pic} has been normalized to the non-magnetic background. Moreover, to improve the signal-to-noise ratio, the intensity of each pixel has been low-pass filtered along the time axis. We have checked that the filtering process does not introduce a significant distortion of the data. For completeness, we compare below the raw and filtered images side-by-side in Figs. \ref{RAWvsFilteredDMI:pic} and \ref{RAWvsFilteredDotCollapse:pic}. Multi-frame movies constructed from the same series of images are included as separate files in the Supplementary Material.
	

\end{document}